\documentclass[preprint2]{emulateapj}
\usepackage{cancel}	
\usepackage{color}

\newcommand{\kms}{km\,s$^{-1}$}

\def\Ha{H{$\alpha$}}
\def\Hb{H{$\beta$}}

\shorttitle{SLSN Ic relation}
\shortauthors{Inserra \& Smartt}

\begin{document}


\title{Erratum: Super Luminous Supernovae as standardizable candles and high redshift distance probes}



\author{C. Inserra and S. J. Smartt}
\affil{Astrophysics Research Centre, School of Mathematics and Physics, Queens University Belfast, Belfast BT7 1NN, UK}
\email{Email : c.inserra@qub.ac.uk}
%







\begin{abstract}
 We investigate the use of type Ic Super Luminous Supernovae (SLSN Ic) as standardizable candles and distance indicators. Their appeal as cosmological probes stems from their remarkable peak luminosities, hot blackbody temperatures and bright rest frame ultra-violet emission. We present a sample of sixteen published SLSN, from redshifts 0.1 to 1.2 and calculate accurate $K-$corrections to determine uniform
magnitudes in two synthetic rest frame filter band-passes with central wavelengths at 400nm and 520nm. At 400nm, we find an encouragingly low scatter in their uncorrected, raw mean magnitudes {\bf with $M(400) = -21.70 \pm 0.47$ mag} for the full sample of sixteen objects. We investigate the correlation between their decline rates and peak magnitude and find that the brighter events appear to decline more slowly. In a similar manner to the Phillips relation for type Ia SNe, we define {\bf a  $\Delta M_{30}$ decline relation}. This correlates peak magnitude and 
decline over {\bf 30} days and can reduce the scatter in standardized peak magnitudes {to $\pm0.25$ mag}. We further show that $M(400)$ appears to have a strong color dependence. Redder objects are fainter and also become redder faster. Using this peak magnitude - color evolution relation, a {\bf low scatter of between $\pm0.19$ mag  and $\pm0.26$ mag} can be found in peak magnitudes, depending on sample selection.  However we caution that only eight to ten objects currently have enough data to test this peak magnitude - color evolution relation. We conclude that SLSN Ic are promising distance indicators in the high redshift Universe in regimes beyond those possible with SNe Ia.  Although the empirical relationships are encouraging, the unknown progenitor systems, how they may evolve with redshift, and the uncertain explosion physics are of some concern. 
The two major measurement uncertainties are the limited numbers of low redshift, well studied, objects available to test these relationships and internal dust extinction in the host galaxies. 
\end{abstract}

\keywords{supernovae: general -- distance scale}

\section{Introduction}\label{sec:intro}

In recent years the new generation of deep, wide surveys like 
the Palomar Transient Factory \citep[PTF,][]{rau09}, the Panoramic Survey Telescope \& Rapid Response
System \citep[Pan-STARRS,][]{PS1_system}, the Catalina Real-time Transient Survey \citep[CRTS,][]{dr09},
and the La Silla QUEST survey \citep{2013PASP..125..683B} have searched for supernovae (SNe) in the local Universe without 
a galaxy bias and have found new types of luminous transients. These were preceded by the smaller scale, but pioneering, Texas Supernova Search \citep{qu05} which  uncovered the first SN
 recognised as ``super-luminous"
\citep{2006CBET..644....1Q,qu07,2007ApJ...666.1116S}.
These super-luminous supernovae (hereafter SLSN) are characterised by absolute magnitudes at
maximum light of $M_{\rm AB} <-21$ mag and total radiated energies of  order $10^{51}$ ergs. 
They are factors of 5 to 100  brighter than type Ia SNe or normal
core-collapse events.  In reviewing the discoveries to date, \citet{gy12} proposed three observational classes for these highly luminous supernovae : SLSN I, SLSN II and  SLSN R based on their photometric evolution and presumed physical characteristics. 

The SLSN I are hydrogen free and show a blue continuum, a distinctive ``W''-shaped spectral feature at $\sim$4200~\AA\/  and a transformation in their spectra to those of normal type Ic SNe. 
They were first recognised by \cite{qu11}, who tied the transients 
SN2005ap and SCP-06F6 \citep{qu07,ba09} together with PTF discoveries by determining their redshifts and common broad
photospheric absorption lines. A detailed study of one of the nearer events (SN2010gx/PTF10cwr at $z=0.23$) by \cite{pa10}
showed that it spectroscopically evolved into a type Ic SN but on much longer timescales than usual. The possibility of quantitatively studying these at much higher redshifts ($z\sim1$) than the known supernova population was shown by  \cite{ch11} in the Pan-STARRS1 Medium Deep field survey. 
\citet{in13} recently reported a detailed study of a sample of five nearby SLSN and explored quantitative models 
to fit the lightcurves. 
They favoured  a model of a  spinning down magnetic neutron star as the source of the energy input to power their luminous light curves. This magnetar model for luminous supernovae
was theoretically proposed by  \cite{kb10}, \cite{2010ApJ...719L.204W} and further investigated by \cite{de12}.  Since the optical spectra of these SNe evolve to be similar to the normal and more common classes of type Ic SNe, \citet{in13} simply termed them SLSN Ic. 

\citet{gy12} suggested a separate definition of 
``SLSN R" mostly based on the observed slow 
lightcurve evolution of SN2007bi \citep{gy09,yo10}
and the fact that this slow decline could be explained
theoretically by the decay of radioactive $^{56}$Ni 
and $^{56}$Co (hence the use of``R"). 
However, recent studies by \citet{ni13} and \citet{mc13} showed that two SLSN which are very similar to SN2007bi are better 
explained by the magnetar model. In fact the physical nature of the radioactive powering is not unambiguously established in any of them which means the label of ``SLSN R" is not one that is observationally secure.  \citet{ni13} 
illustrated the similarity in the early spectra of the SLSN I types  \citep[of][]{pa10,qu11,in13} 
and the SN2007bi-like events (PTF12dam in 
particular). It is certainly possible the ``SLSN I" and the ``SLSN R" explosions have the same physical 
mechanism. However, it is not clear yet if there is a continuum in their decay timescales or if there are two 
distinct groups which decline with different rates after peak. Following \cite{in13} and \cite{ni13}, we will use the classification term SLSN Ic to refer to {\em all} the 
hydrogen poor super-luminous supernovae. Their physical nature may be debated, but the simple observational term of type Ic is quite secure in the classical scheme of SN classification. In order to distinguish between the two observational groups 
we will use the terms SN2005ap-like and SN2007bi-like events since these are the prototypes of the faster and slower evolving events known to date. 
 
The SLSN of type II generally show strong H\,{\sc i} lines in emission which indicate the interaction of 
fast moving ejecta with a pre-existing and slower moving shell of material. In this scenario, the kinetic energy of the 
ejecta (with energies of order 10$^{51}$ergs) is converted to luminosity during the collision 
\cite[e.g.][]{2007ApJ...671L..17S,2009ApJ...691.1348A}. These have been relatively easy to identify due to the
strong emission line signatures, although there is still uncertain physics concerning the symmetry and clumping of the interacting shells. 
The interaction model has been extended to explain the hydrogen free SLSN by 
\cite{2011ApJ...729L...6C} 
and 
\cite{2012ApJ...757..178G}. 
\cite{be14} have recently shown that the 
transient CSS121015:004244+132827 appears quite similar to many of the SLSN Ic of \cite{qu11} and \cite{in13} but weak and transitory H$\alpha$ emission is detected, suggestive of interaction occurring. 

The number of  SLSN Ic which have been discovered and recognised as such has increased rapidly, 
with discoveries out to redshifts of $z\sim4$ \citep{ch11,co12,be12,le12,in12a,lu13,ho13,ko13,be14,mc14b}. They seem to be fairly homogeneous in their spectroscopic and 
photometric properties and have attracted considerable attention due to their potential utility as cosmological standard candles. The potential lies in the fact that they are 2-3 magnitudes brighter than type Ia SNe and their
hot blackbody temperatures mean that their UV continuum emission is detectable at high redshifts in the optical and near-infrared \cite[e.g][]{co12}.  The study of these objects is still a relatively youthful endeavor as they were
first recognised as highly luminous in 2010-11 \citep{pa10,qu11,ch11}.  \citet{qu13b} already pointed out that the
peak magnitudes of  SLSN Ic  may be quite narrowly distributed at 
$M = -21.7 \pm 0.4$ (in the the unfiltered ROTSE-IIIb  band-pass). Detailed investigation of this issue is now
possible with a large sample of well observed SLSN Ic, which have multi-color light-curves and spectral
sequences to allow more accurate $K$-corrections to be determined.  \cite{ki13} 
discussed how novel high redshift standard candles could be used to probe the dark energy
equation of state $w(z)$ and how this compares to the current state-of-the-art type Ia supernova and
baryon acoustic oscillation measurements. It would appear that a large number ($\sim10^{3}$) are required to improve on $w(z)$ estimates, and objects should cover a long redshift baseline, which is challenging. Nevertheless, any additional cosmological probes  are potentially useful and their merits should be studied.

In this paper we analyse a sample of well observed SLSN Ic with complete light-curves from before peak 
to beyond 30 days.  We test whether there is any relation between empirical, observable 
quantities (such as light-curve shape, color evolution and peak luminosity) that could make them 
useful cosmological standardizable candles. It is well known that absolute magnitudes of type Ia SNe have been standardized using different methods based on their empirical characteristics. 
The earliest and simplest standardization method correlated peak magnitudes with decline rates and is referred to as the $\Delta m_{15}$ method \citep{ru74,ps77,ph93,ham96}. This gave way to much more sophisticated approaches such as 
 the Multicolor Light Curve Shapes method \citep[MLCS,][]{rie96,rie98}, the stretch method \citep{pe97,go01}, the spectral adaptive light curve template method \citep[SALT,][]{guy05,guy07} and the BayeSN method \citep{man09,man11}.
In the context of this initial attempt to standardize SLSN Ic, it is useful to reflect on the benchmark results from the
early attempts with the $\Delta m_{15}$ method. \cite{ph93} used 9 SNe and illustrated the large dispersion in the raw 
absolute magnitudes (without any correction) of $\sigma=0.8$, 0.6 and 0.5 for $B,V$ and $I$ respectively 
\footnote{The $\sigma$ for the raw absolute magnitudes is simply the standard deviation of the population, as calculated by \cite{ph93}. The $\sigma$ or scatter around the best fit line for $\Delta m_{15}$ is the rms error of the linear fit.}.
The  Phillips $\Delta m_{15}$ correlation immediately reduced the scatter to  $\sigma=0.36$, 0.28 and 0.38 in $BVI$. 
This method was revisited with a much larger sample by \cite{pr06}, who produced an rms scatter around the linear fit of   
$\sigma=0.17, 0.14$ for 94 SNe in $BV$ and $0.15$ for 87 SNe in the $I$-band. Since that time, the community has 
been investigating other standardizable candles with a view to reaching similar scatter values. Type II-P SNe have been 
proposed, which appear to have an rms scatter in their standardizable luminosity-velocity relation of around 0.26 mag 
\citep{2002ApJ...566L..63H,2006ApJ...645..841N,2009ApJ...694.1067P,2010ApJ...708..661D}. 
\cite{2010MNRAS.403L..11M} suggested that going to the near infrared (NIR) could potentially reduce this to 0.1-0.15 mag, although the disadvantage of SNe II-P as distance indicators is that they are 2-3 magnitudes fainter than SNe Ia. We note that also GRB have been suggested to be possible standardizable candle \citep{2014arXiv1407.2589C,2014arXiv1409.3570C}. These attempts illustrate that achieving an rms scatter of 
$0.1 - 0.3$ magnitudes for any new standardizable candle would  make it competitive and interesting to study further. This paper investigates the use of SLSN Ic as such standardizable candles.

\begin{deluxetable*}{lccccccccccc}
\tablewidth{0pt}
\tablecaption{Sample of SLSN.\label{table:data}}
\tablehead{
\colhead{SN} & \colhead{$z$} &\colhead{E(B-V)} & \colhead{$m$}& \colhead{Ref.$^a$} & \colhead{Filters 400,520} & \colhead{$A_{f}$} 
& \colhead{$M(400)$} & \colhead{$\Delta M_{10}$(400)} & \colhead{$\Delta M_{20}$(400)} & \colhead{$\Delta M_{30}$(400)} & \colhead{$\Delta$M(400-520)$_{30}$}
}
\startdata
\cutinhead{SLSN Ic (2005ap-like)}
SN2011ke&	0.143&	0.01&	17.70~(g)&1	&g$\rightarrow 400$, r$\rightarrow 520$&	0.05	&-21.27&	0.46&	1.22	&2.14	&0.65\\
SN2012il	&      0.175	&0.02&	18.34~(g)*&1	&g$\rightarrow 400$, r$\rightarrow 520$&	0.09&	-21.31&	0.41	&1.06&	1.74	&0.50\\
PTF11rks	&      0.190	&0.04&	19.13~(g)&1	&g$\rightarrow 400$, r$\rightarrow 520$&	0.17	&-20.63&	0.28	&0.95&	2.02&	0.95\\
SN2010gx&	0.230&	0.04&	18.43~(g)&2	&g$\rightarrow 400$, r$\rightarrow 520$&	0.13&	-21.73&	0.13&	0.63&	1.41&	0.58\\
SN2011kf&	0.245&	0.02&	18.60~(g)&1	&g$\rightarrow 400$&	0.09&	-21.76&	0.09&	0.54&	1.25&	-\\
LSQ12dlf&	0.255&	0.01&	18.78~(V)&3	&V$\rightarrow 400$, R$\rightarrow 520$&	0.03	&-21.59&	0.38	&0.73&	1.34&	0.55\\
PTF09cnd&	0.258&	0.02&	18.09~(R)&4	&R$\rightarrow 400$&	0.05	&-22.15&	0.19&	0.50&	0.94&	-\\
SN2013dg&	0.265&	0.04&	19.26~(g)*&3	&g$\rightarrow 400$, r$\rightarrow 520$&	0.15	&-21.38&	0.43	&1.24&	2.00&	0.75\\
SN2005ap&	0.283&	0.01&	18.35~(R)&5	&R$\rightarrow 400$&	0.02	&-22.12&	0.19&	0.66&	-&	-\\
PS1-10bzj&	0.650&	0.01&	21.44~(r)*&6	&r$\rightarrow 400$, z$\rightarrow 520$&	0.02	&-21.11&	0.21	&0.81&	1.55&	0.66\\
PS1-10ky	&      0.956&	0.03&	21.27~(i)&7	&i$\rightarrow 400$, z$\rightarrow 520$&	0.06&	-22.07&	0.21&	0.90&	1.31&	0.31\\
SCP-06F6&	1.189&	0.01&	21.04~(z)&8	&z$\rightarrow 400$&	0.01	&-22.17&	0.11	&0.44&	0.90&	-\\
PS1-10pm&	1.206&	0.02&	21.74~(z)&9	&z$\rightarrow 400$&	0.02	&-22.03&	0.15	&0.43&	-&	-\\

\cutinhead{SLSN Ic (2007bi-like) }
PTF12dam&0.107&0.01& 16.84~(g)* & 10&g$\rightarrow 400$, r$\rightarrow 520$& 0.04 & -21.56 & 0.02 & 0.15 & 0.28&  0.08\\
PS1-11ap&0.524&0.01&20.11~(r) &11 &r$\rightarrow 400$, i$\rightarrow 520$& 0.02& -21.82 & 0.09 & 0.28 & 0.60 & 0.11\\
\cutinhead{SLSN II}
CSS121015 &0.287&0.08& 18.19~(V) & 12&V$\rightarrow 400$, R$\rightarrow 520$& 0.24 & -22.48  & 0.25 & 0.76 & 0.93 &  0.21
 \enddata
 \tablecomments{From left to right :  SN designation; measured redshift; foreground extinction; observed magnitude (AB system) at peak with the band used in parentheses; reference; observed filters used to calculate the synthetic 400nm and 520nm magnitudes; extinction toward the SN in the observed band; absolute magnitude in the 400nm band,  magnitude decrease in 10, 20 and 30 days; the color change between the 400nm and 520nm synthetic bands at peak and 30 days after.\\
$^a$ References: 1. \citet{in13}, 2. \citet{pa10}, 3. \citet{ni14}, 4. \citet{qu11}, 5. \citet{qu07}, 6. \citet{lu13}, 7. \citet{ch11}, 8. \cite{ba09}, 9. \citet{mc14b}, 10. \citet{ni13},  11. \citet{mc13}, 12. \citet{be14}.\\
$^*$ The published magnitude closest in time to the peak of the polynomial fit.}
\end{deluxetable*}

\section{The  sample and data }\label{sec:sample}

The sample chosen was selected from the published SLSN Ic  which have well sampled light-curves 
around peak luminosity. We required photometric coverage from several days pre-maximum (or a good estimate of the time and value of peak luminosity) to about 35 rest frame days  after peak luminosity. We further required
photometry in observer frame filters which, after $K$-correction, had similar equivalent rest frame bandpasses (see Section\,\ref{sec:rel}).  An accurate spectroscopic redshift is also required to provide reliable relative distances.
The redshift distribution of the known population of  well observed SLSN ($0.1 \leq z \leq 1.6$) 
means this selection is not trivial and not all the 30-40 or so known SLSN can be used in our sample calculations 
\citep[see][for an extensive, but not complete, compilation of 31 SLSN]{lu13b}. 
We found sixteen SNe which had suitable data for our analysis:  SN2010gx, SN2011kf, SN2011ke, SN2012il, SN2005ap, PS1-10ky, PTF09cnd, SCP-06F6, PTF11rks, PS1-10bzj, PS1-10pm, LSQ12dlf, SN2013dg, PS1-11ap, PTF12dam and CSS121015:004244+132827
 \citep[with data from][]{pa10,in13,qu07,ba09,qu11,ch11,lu13,ni13,mc13,be14,mc14b,ni14}. The details are listed in Table\,\ref{table:data}.
We highlight the important caveat  that this sample selection does not take potential Malmquist bias into account. The identification process of these targets comes from a range of 
surveys, each with quite different magnitude limits combined with different selection procedures
for selecting targets for spectroscopic follow-up. Hence the process from discovery of a transient to 
identification as a super-luminous supernova is quite  inhomogeneous.  It is certainly true 
that the intrinsically brightest SNe are over represented in magnitude limited surveys 
compared to their volumetric rates which may introduce a shift in absolute magnitude 
simply because the lower luminosity events are under-represented. There does appear
to be a significant gap between the SLSN Ic and the brightest (and most energetic) of 
type Ic SNe \citep[e.g. see][for a comparison]{in13}.
 Whether this gap is Malmquist type bias or real remains to be investigated.

\begin{figure}
\center
\includegraphics[width=\columnwidth]{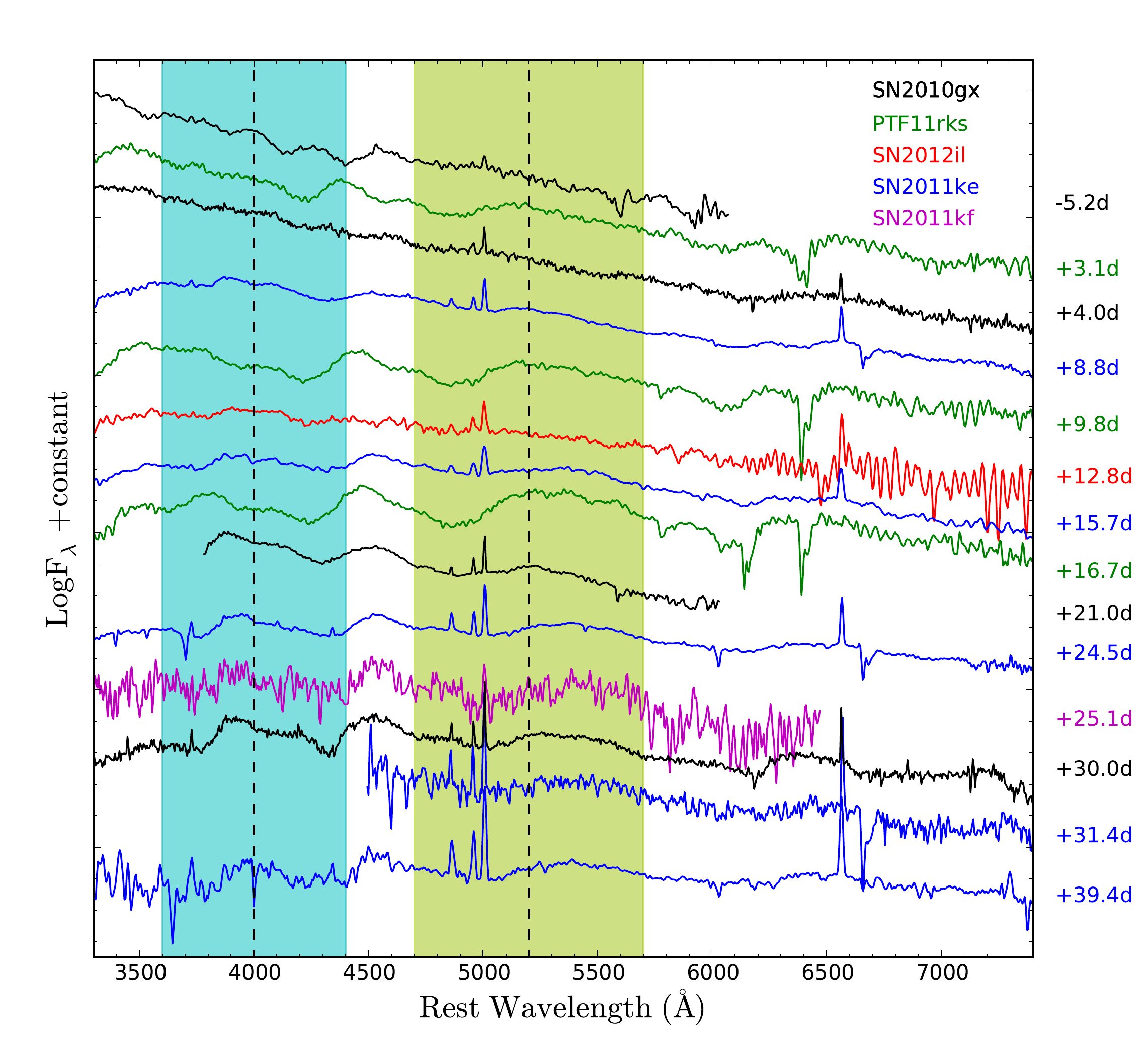}
\caption{General spectral evolution of SLSN Ic (2005ap-like) from 5d before maximum light to 40d after. 
The phase of each spectrum relative to light curve peak in the rest frame is shown on the right. These restframe spectra have been corrected for Galactic extinction. They have been convolved with a Gaussian function of FWHM = 5~\AA\/  and subsequently binned to 5~\AA\/ per pixel.
The 400nm and  520nm wavebands are shown in cyan and green respectively, with their effective
centers marked with black dashed lines. 
The data are  from \citet{pa10} and \cite{in13}. Table\,1 provides a full list of references to the original sources of data for all objects. }
\label{fig:comp}
\end{figure}

Thirteen of these objects appear to be similar to the well observed SN2010gx, and these decay rapidly after peak magnitude. The similarity of their lightcurves and spectral evolution has been discussed by \cite{qu11}, \cite{in13} and \cite{gy12}, amongst others and we label these 2005ap-like since that was the first one discovered. 
Two of our sample of sixteen  (PTF12dam and PS1-11ap) decline on noticeably slower timescales and are similar to SN2007bi (see Section~\ref{sec:intro}). 
We include them in our analysis for completeness and show results
with and without these two events (and refer to them as 2007bi-like). 
The object CSS121015:004244+132827 \citep[hereafter CSS121015,][]{be14}, 
has been classified as a SLSN II since it shows signs of H$\alpha$ that is probably linked to the supernova rather than the 
host galaxy. We include it here for completeness, since its spectra resembles SN2010gx and the other SLSN Ic with the 
addition of weak hydrogen features, and at high redshift one might struggle to see the subtle differences present in spectra. 
Again, we will show results with and without this event. We note that although we included it and label it a SLSN II, we
are not considering the broader and more populous class of SLSN II, such as the SN2006gy-like events. These luminous type IIn SNe are unambiguously driven by strong CSM interaction and collisions of hydrogen rich gas and they are observationally very distinct. 
Hereafter we will refer to three samples. The thirteen 2005ap-like  events are called the ``pure sample". Including the two 2007bi-like objects makes an ``extended sample" of fifteen objects. And adding in CSS121015 defines the  ``complete sample" of sixteen objects.

We included all  SLSN having enough data to be useful for our analysis.
The SLSN published in the literature which are  
not reported and used here were rejected primarily on the basis that they lack the photometric data required for our careful and uniform analysis. 
There are at least seven other SLSN Ic which are securely classified in previous work and have data published, but their photometric coverage 
does not present a complete enough dataset around the epoch of 0-30d (rest-fame) in filters that are suitable
for the comparison we undertake in Section\,\ref{sec:rel}. 
These are SN2006oz \citep{le12},  PTF09cwl and PTF09atu \citep{qu11}, PTF10hgi \citep{in13}, PS1-10awh \citep{ch11}, PS1-11bam \citep{be12} and SSS120810:231802-560926 \citep{ni14}.  
We did not consider PS1-10afx further because of the probable nature of it being a lensed type Ia SN and not a SLSN  \citep[][]{chor13,qu13a,qu14}.
We also did not include the slowly fading SLSN SN2007bi \citep{gy09,yo10}, as it lacks extensive photometric coverage around peak in the blue bands. SN2007bi  is not as well sampled in standard filters compared to the excellent coverage of the
phenomenologically similar events PTF12dam and PS1-11ap \citep{ni13,mc13}.

In our measurements of the absolute magnitude we considered only the reddening values for interstellar dust in our Galaxy (see Table~\ref{table:data}). The SNe of the sample were located in faint, dwarf galaxies and were unlikely to have suffered significant additional extinction. 
Indeed no absorption due to Na\,{\sc I} D  lines from gas in the hosts was observed in any spectrum.
This does not meant that  we can exclude {\it a priori} possible dust extinction from the host. The assumption that host extinction is uniformly negligible is probably the biggest uncertainty in using SLSN Ic as standard candles.  
In all cases a standard reddening curve was assumed with $R_{V}=A_{V}/E(B-V)=3.1$.
The objects cover a range of redshift between 0.143 and 1.206, hence in the next section we discuss the appropriate corrections needed for redshift effects ({\it K}-correction and time dilation) to obtain the absolute rest frame peak magnitudes.

\section{Data Analysis}\label{sec:da}

\subsection{Redshift}\label{sec:red}
We assume a cosmology with $H_{0}=72$ \kms\/ Mpc$^{-1}$, $\Omega_{m}=0.27$ and $\Omega_{\Lambda}=0.73$.
All the redshifts used in this paper were evaluated spectroscopically from  host galaxy lines (either emission,  Mg\,{\sc ii} absorption or both) with the exceptions of LSQ12dlf and SN2013dg.  
For these two, cross-correlation with a library of SLSN spectra was employed, retrieving a typical error on the redshift of $\pm0.005$ \citep[see][for details]{ni14}. 
In the cases of SN2010gx \citep{pa10,qu11}, SN2005ap \citep{qu11}, PS1-10ky \citep{ch11}, PTF09cnd \citep{qu11}, SCP06F6 \citep{qu11}, PS1-10bzj \citep{lu13} and PTF11rks \citep{qu11b} we retrieved the redshifts by analysing the narrow absorption lines of Mg~{\sc ii} doublet $\lambda\lambda$2796, 2803 and compared to those previously published.
For the other three SLSN Ic, we evaluated the redshifts by the identification of host galaxy lines, such as [O {\sc ii}] $\lambda$3727, [O {\sc iii}] $\lambda\lambda$4959, 5006, \Ha\/ and \Hb\/ \citep[cf. with those of][]{qu10,dr11,qu11c}. In the case of SN2010gx  the redshift evaluated by the host galaxy lines \citep{pr12,pa10b,pa10} is identical to that retrieved through the analysis of Mg~{\sc ii} lines \citep{qu10,qu11}, and there are no cases where there is any measurable difference between 
the two methods. 
The two 2007bi-like events (PTF12dam and PS1-11ap) show Mg~{\sc ii} lines in absorption. Again, the redshifts  determined from this doublet were checked with the emission lines of the host galaxies \citep[as previously done by][]{ni13,mc13}. 
The typical error on the spectroscopic redshift is $\pm0.003$, which corresponds to uncertainties in distance moduli of 
0.07, 0.02 and 0.01 at redshifts of $z=0.1, 0.5$ and 1.0 respectively.

\begin{figure}
\includegraphics[width=\columnwidth]{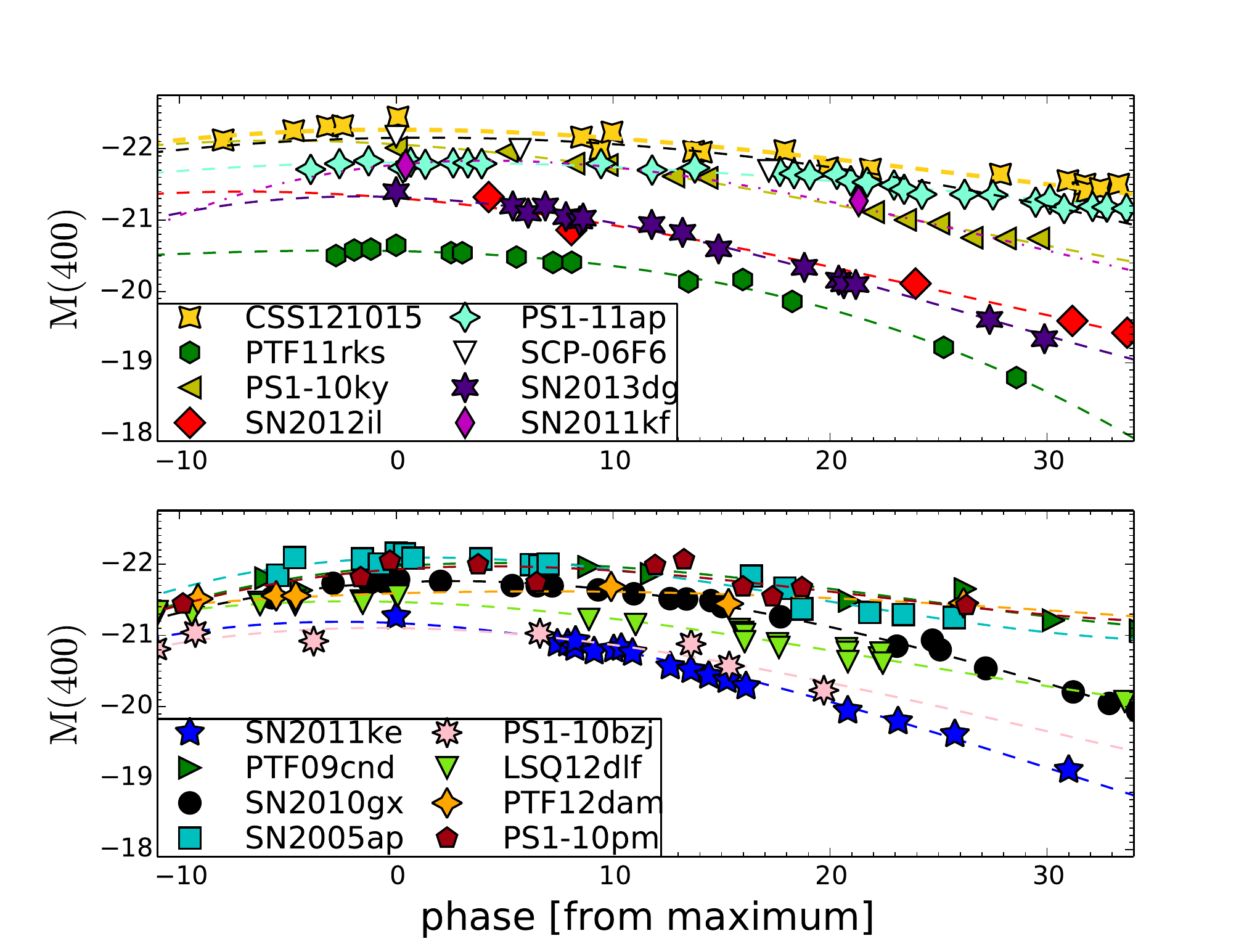}
\caption{Light curves of our SLSN Ic sample listed in Table~\ref{table:data}. The absolute magnitudes, $M(400)$ are  for the 400nm band after $K$-correction. Time dilation has been applied to report he phase in restframe days with respect to maximum light. Polynomial fits to the photometry of the sixteen SNe are  shown as broken lines.  The line fits are all  third order polynomials, apart from SN2011kf and SCP06F6 (second order polynomials) and CSS121015 (fourth order polynomial). The objects are split into two panels simply for clarity and illustration purposes.} 
\label{fig:lcfit}
\end{figure}

\subsection{$K$-corrections and synthetic photometric bands}\label{ss:kc}
Photometric and time dependent measurements of objects at significant cosmological redshifts
require application of the $K$-correction and time dilation correction in order to provide meaningful comparative rest frame properties. 
The $K$-correction is essential because measurements through a filter in the observer frame 
detect photons from a different part of the rest frame spectrum than the filter throughput
\citep[e.g.][]{1968ApJ...154...21O,2002astro.ph.10394H,br07}. This means that an accurate determination
for the $K$-correction requires a flux calibrated spectral energy distribution (SED) which covers
a wavelength range large enough for the correction to be calculated. This is ideally done with 
spectra but  multi-color photometric measurements can also be usefully employed to constrain the 
observed SED. Such $K$-corrections have been calculated since the early days of using type Ia
SNe as distance indicators \cite[e.g.][]{1993PASP..105..787H,1996PASP..108..190K,nu02}
and more recently large libraries of SN Ia spectra have been employed for detailed determination 
\citep{2007ApJ...663.1187H}.   At low redshift, it is relatively straightforward to calculate corrections
to observer frame filter magnitudes such that the rest frame magnitudes are in the same filter band pass. 
However, at redshifts beyond $z \gtrsim 0.2$, the de-redshifted  observed filter function is often closer
to another, bluer, standard filter and hence it is more useful to determine $K$-corrections to a different 
filter than observed (``cross-filter" K-corrections). For example, at $z=0.28$ the effective central wavelength of the 
Pan-STARRS1 $r_{\rm P1}$ filter corresponds to the rest frame $g_{\rm P1}$ filter and hence a 
$K$-correction to correct the observed $r_{\rm P1}$ to rest frame $g_{\rm P1}$   is useful
\citep[see][for filter throughput details]{2012ApJ...750...99T}. 
When the observed and rest frame filter bands are significantly 
misaligned, the $K$-corrections are of course more dependent on the assumed spectral template.  The
corrections can be computed by using either a theoretical or empirical template spectrum (or SED)
at the relevant epoch. Interpolations of spectra either side of the particular photometric epoch 
can be used. \citet{nu02} suggested that for SN Ia the dispersion in the cross-filter $K$-corrections
is less than 1\% when the interpolated spectra are constrained within $\pm$4\,days of the epoch, whilst 
\citet{2007ApJ...663.1187H} showed that with a similar baseline the scatter is probably higher, at 5\%. 

Typically, $K$-corrections are calculated to provide magnitudes in rest frame filters that are 
in a standard system, such as SDSS $ugriz$, 
or Johnson-Cousins $UBVRI$. 
In order to make a meaningful comparison of the SLSN Ic absolute magnitudes we
need to correct their observed magnitudes to a single filter.  While it would be possible
to correct to a standard filter in a standard system, it is not clear that this is particularly
useful for our purposes. If SLSN Ic are standardizable candles, then their future use is 
likely to be in redshift regimes beyond those probed by SNe Ia. There is little 
to be gained by correcting to currently defined rest frame filters since the lack of 
low redshift SLSN Ic, which have negligible $K$-corrections, renders the  standard filters fairly 
useless\footnote{This lack of objects known below $z<0.1$ is simply due 
to their low volumetric rates.}. Instead we choose to define two synthetic rest frame filter band-passes to correct to. 

Analysis of the SN rest frame spectra during the first month from the maximum light \citep{pa10,qu11,ch11,in13}, indicates that the region around 4000 \AA\/ is continuum dominated and relatively free from strong spectral features that could increase the photometric dispersion from object to object. 
To determine synthetic magnitudes at this rest-wavelength window, we built the synthetic passband with an effective width of 800 \AA\/ and central wavelength at 4000 \AA\/, having steep wings and a flat top (more similar to Sloan than Bessel filters to decrease the errors in the $K$-correction process). Hereafter we refer to it as the 400 nm band. 
In a similar way, we defined a second band peaking at 5200 \AA\/ with an effective width of 1000 \AA\/, and hereafter refer
to it as the 520nm band. Fig.\,\ref{fig:comp} illustrates these synthetic passbands and the rest frame spectral
regions lying in them. 
The choice of these bands was motivated by both the lack of strong spectral features and 
also by several practical considerations.  To maximise the sample size, we required appropriate photometric data that could be $K-$corrected uniformly to these two bands.
If we move the primary restframe band (400nm band) redder  (for example close
to the standard SDSS $r-$band which is also relatively featureless) we 
reduce the complete sample from sixteen objects to eleven. This is due to a lack of photometric
data at the appropriate redshifted wavelength region. Furthermore we would not then be able to 
define a second band (for color tests) since a restframe SDSS $i$-band (or something similar) 
would lie beyond the observers $z-$band, in the near-infrared (NIR). We lack any substantial data for these objects in the  NIR.   One could consider going the other way and pushing the
primary 400nm band bluer, but one must be careful to avoid the strong (and potentially diverse) absorption associated with elements such as Mg~{\sc ii} and Si~{\sc iii} \citep[see][]{ch11}. We did identify a potentially useful region around 3000~\AA, which is the bluest we could consider given the 
lack of extensive restframe UV spectroscopy. However, using this would reduce our complete sample to just five objects because of a paucity of data in the observer $u$- or U-band for objects with 
$z\lesssim0.3$. 

For each SN in question, we calculated $K$-corrections from a suitable observer
frame filter back to the 400nm and 520nm bands. The filters chosen were those that  had 
effective central wavelengths most closely matched  to $400(1+z)$ and $520(1+z)$\,nm and are listed in Table\,\ref{table:data}. These are typically $g$ and $r$-band but the higher redshift objects employ $i$ and $z$-bands as appropriate. For five of the objects, the available photometry only allowed correction to the bluer 400nm band. 
When the observed filters used were Bessel, we converted the Vega magnitude into the AB system using the prescription presented in \citet{br07}.

\begin{figure}
\includegraphics[width=\columnwidth]{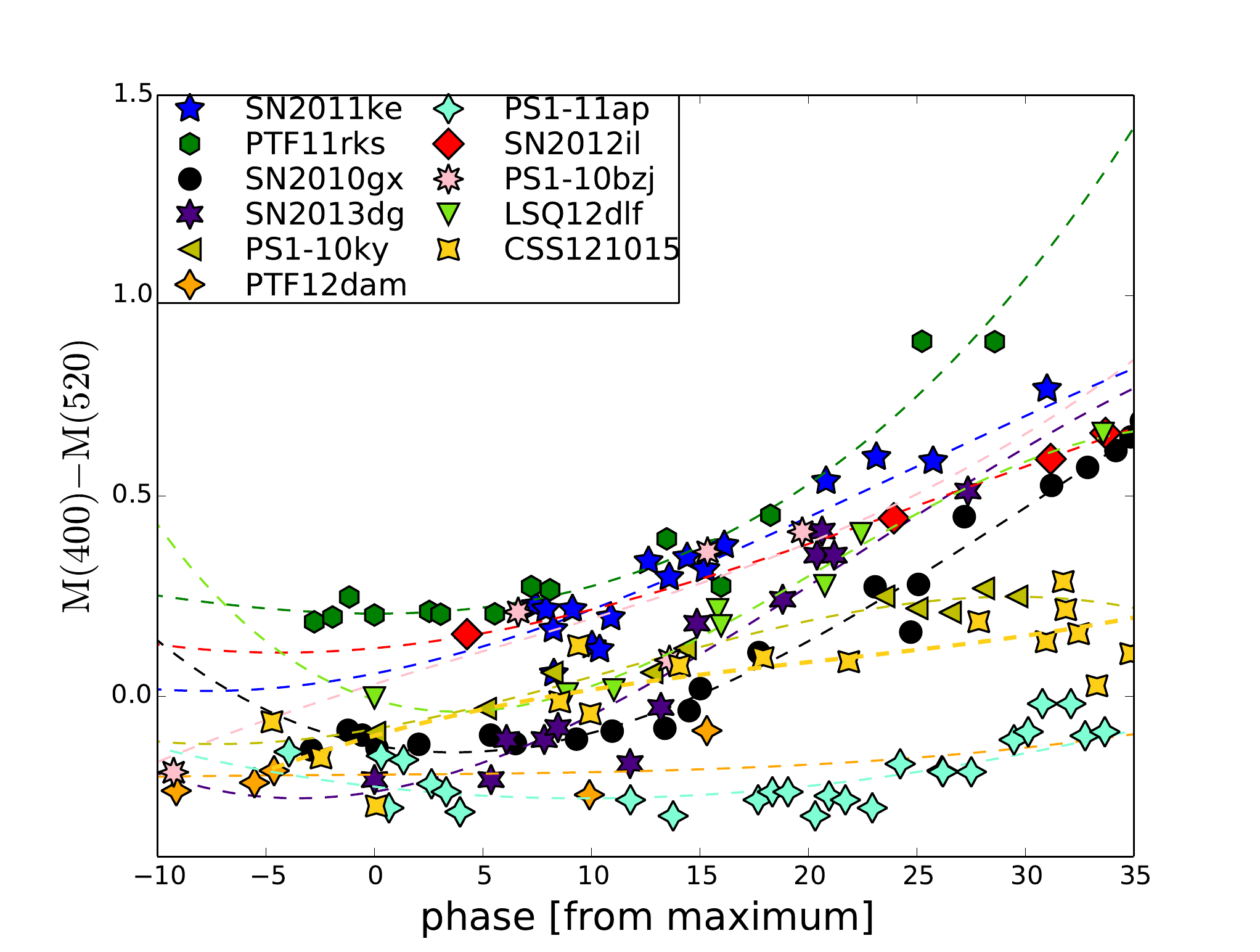}
\caption{Color evolution as measured from the $M(400) - M(520)$ color index vs restframe phase 
in days. The reference phase of zero days is the epoch of maximum in the 
400nm band. $K$-corrections and time dilation have been applied. Polynomial fits to the color evolution are shown. The color of SN 2012il close to peak was evaluated using spectra (see Appendix\,\ref{app:fits} 
for details).  All line fits represent third order polynomials (dashed lines), apart from that for 
PS1-10ky which was fitted with a fourth order polynomial (dotted line).} 
\label{fig:colfit}
\end{figure}

We computed $K$-corrections with our own $python$ based code which we term SuperNova Algorithm for $K$-correction Evaluation (SNAKE; a full description will be published in Inserra et al. in preparation along with a code release).
We calculated $K$-corrections for each photometric point available in the literature for each SN using 
an observed spectrum at the same, or similar epoch within $\pm4$ days from the photometry. For photometric epochs for which no spectra were available, we determined a SED using the multi-color photometric measurements available. If all colors were not available at the specified epoch, we 
allowed a window of $\pm2$ days to select the photometric measurements. If both spectra and photometry were absent we used our library of spectra that includes 78 spectra from seventeen SLSN Ic (including 2007bi-like SLSN).  In Table~\ref{table:kcorr} we report the $K$-corrections at key epochs along with the method used to evaluate them (see~Appendix~\ref{app:err} for error treatment and differences between the three methods). A time dilation correction of $(1+z)$ was applied to all epochs with respect to the peak magnitude which was defined as $phase = 0$.  The absolute magnitudes were then calculated from 

\begin{equation}
m_{f} = M(400) + \mu  + K_{f \rightarrow 400} + A_{f}\,, 
\label{eqn:absmag}
\end{equation}

where $m_{f}$ is the AB magnitude in the observed filter $f$,  $\mu$ is the distance modulus calculated from the luminosity distance for our adopted  cosmology of $H_{0}=72$ \kms\/ Mpc$^{-1}$, $\Omega_{m}=0.27$ and $\Omega_{\Lambda}=0.73$; 
$A_{f}$ is the Galactic extinction in the observed filter ; and $K_{f \rightarrow 400}$ is the 
$K$-correction from the observed filter in Table\,\ref{table:data} to the synthetic 400nm bandpass. 
The $M(400)$  lightcurves and the $M(400) - M(520)$  color evolution curves of all the SNe after these $K$-corrections are shown in Fig.~\ref{fig:lcfit} and Fig.\,\ref{fig:colfit}. 
The errors on the absolute magnitudes are estimated by propagating the errors of the four
terms in Equation\,\ref{eqn:absmag}, including an estimate of the uncertainty in the $K-$correction (details are in Appendix\,\ref{app:err}).

\begin{deluxetable*}{lccccccc}
\tablewidth{0pt}
\tablecaption{Calculated $K$-corrections for the sample of SLSN.\label{table:kcorr}}
\tablehead{
\colhead{SN} & $z$ & \colhead{$K_{f \rightarrow 400}^{\rm peak}$} & \colhead{$K_{f \rightarrow 400}^{\rm 10}$} & \colhead{$K_{f \rightarrow 400}^{\rm 20}$} & \colhead{$K_{f \rightarrow 400}^{\rm 30}$} & \colhead{$K_{f \rightarrow 520}^{\rm peak}$} & \colhead{$K_{f \rightarrow 520}^{\rm 30}$}
}
\startdata
\cutinhead{SLSN Ic (2005ap-like)}
SN2011ke& 0.143&	-0.18 (P)&	-0.17 (S)&	-0.22 (S)&	-0.24 (P) &-0.11 (S)&-0.16 (S)\\
SN2012il& 0.175&	-0.18 (L)&	-0.18 (S)&	-0.23 (L)&	-0.25 (P) &-0.09 (L) &-0.15 (P)\\
PTF11rks& 0.190&	-0.19 (S)&	-0.17 (S)&	-0.16 (S)&	-0.17 (P)&-0.19 (S)&-0.19 (P)\\
SN2010gx &0.230 &	-0.23 (S) & -0.18 (S)&	-0.18 (S)&	-0.07 (S)&-0.23 (S)& -0.20 (S)\\
SN2011kf& 0.245&	-0.13 (L) & -0.10 (L) &	-0.08 (S)&	-0.08 (S) &-&-\\
LSQ12dlf& 0.255 &  -0.17 (S) & -0.15 (S) & -0.29 (S) & -0.38 (S) & -0.28 (S) & -0.20 (S)\\
PTF09cnd&0.258	&      -0.34 (S)&	-0.29 (L)&	-0.29 (L)&	-0.16 (L)&-&-\\
SN2013dg& 0.265& -0.30 (P) & -0.25 (S) & -0.40 (S) & -0.47 (S) & -0.45 (S) & -0.40 (S)\\
SN2005ap& 0.283&	-0.33 (S)&	-0.57 (L)&	-0.60 (L)&	- &-&-\\
PS1-10bzj&0.650 &	-0.58 (P) &	-0.57 (S)&	-0.46 (S)&	-0.48 (L) &-0.37 (P) &-0.53 (L)\\
PS1-10ky& 0.956&	-0.73 (S)&	-0.75 (P)&	-0.76 (P)&	-0.68 (S) &-0.65 (P)&-0.59 (P)\\
SCP-06F6& 1.189&	-1.35 (L)&	-1.45 (L)&	-1.51 (L) &	-1.56 (L)&-&-\\
PS1-10pm&1.206 &	-0.84 (P)&	-0.96 (P)&	-1.04 (P)&	-&-&-\\

\cutinhead{SLSN Ic (2007bi-like)}
PTF12dam& 0.107& -0.07 (S) & -0.02  (S) &+0.04 (S) & +0.04 (S) &-0.03 (S)&+0.01 (S)\\
PS1-11ap& 0.524&  -0.45 (S) & -0.42 (S) & -0.43 (P) & -0.47 (P) & -0.47 (S)& -0.53 (P)\\
\cutinhead{SLSN II}
CSS121015& 0.287 & -0.37 (S) & -0.36 (S) & -0.35 (S)& -0.32 (P) &-0.40 (S)&-0.22 (P)
\enddata
\tablecomments{(S) denotes that a spectrum of the same object has been used to evaluate the $K$-correction. (P) indicates that the multi-color photometry of the object was used, while (L) means that we used our library of spectra.}
\end{deluxetable*}

\subsection{Absolute peak magnitudes and light-curve interpolation}\label{sec:int}

Most of the sample of sixteen SNe had enough data points around peak to secure a 
confident estimate of both the value and epoch of the peak magnitude in the 400nm band. 
However not all SNe had specific measurements exactly at the four key epochs we identified 
as having useful diagnostic power. These epochs are at peak (defined as 0 days) and 
after +10, +20 and +30 days (rest frame). We employed a consistent method 
of polynomial fits to the $M(400)$ data points to allow estimates of the magnitudes at these epochs. 
We found polynomials of order from 2 to 4 were sufficient and the results of these fits are 
plotted over the data points in Fig.\,\ref{fig:lcfit}. Similarly, polynomial fits to the
 $M(400)-M(520)$ color curves were calculated (either third or fourth order) and are plotted  
in Fig.\,\ref{fig:colfit}.  The specific details for each light-curve fit are given in the Appendix~\ref{app:fits}.
The fits displayed in Figures~\ref{fig:lcfit}~\&~\ref{fig:colfit}  and the magnitudes reported in Table~\ref{table:data} are those from the 
values of the polynomial interpolation at the specific epochs of peak, +10d, +20d and +30d, unless otherwise noted (see Appendix~\ref{app:fits} for details on each object). 

The peak absolute magnitudes of the pure sample (see Section~\ref{sec:sample}) 
give a mean 
{\bf of $M(400) = -21.64 \pm 0.46$ where the latter number is the standard deviation of the sample\footnote{To be clear, we are assuming that these thirteen objects are a sample of a larger population, hence the sample standard deviation is calculated using $(n-1)$ as a denominator. }. 
If the two slowly declining, 2007bi-like SLSN are included then the extended sample mean and standard
deviation are not significantly changed $M(400) = -21.65 \pm 0.43$. 
We also obtain a similar scatter of $M(400) = -21.70 \pm 0.46$ in the complete case.} 
 This raw, and uncorrected range is quite encouraging since the bulk of normal (often historically known as ``Branch" normal) type Ia SN 
originally had uncorrected peak magnitudes with scatters of $\pm$0.8, 0.6 and 0.5 in $BVI$ respectively
\citep{ph93}.  The sample of \cite{pr06} was a factor of ten larger and showed slightly higher values
for the raw standard deviation  $\pm$0.9, $0.7$, $0.6$, $0.6$ in $BVRI$ respectively.

\begin{figure}
\includegraphics[width=\columnwidth]{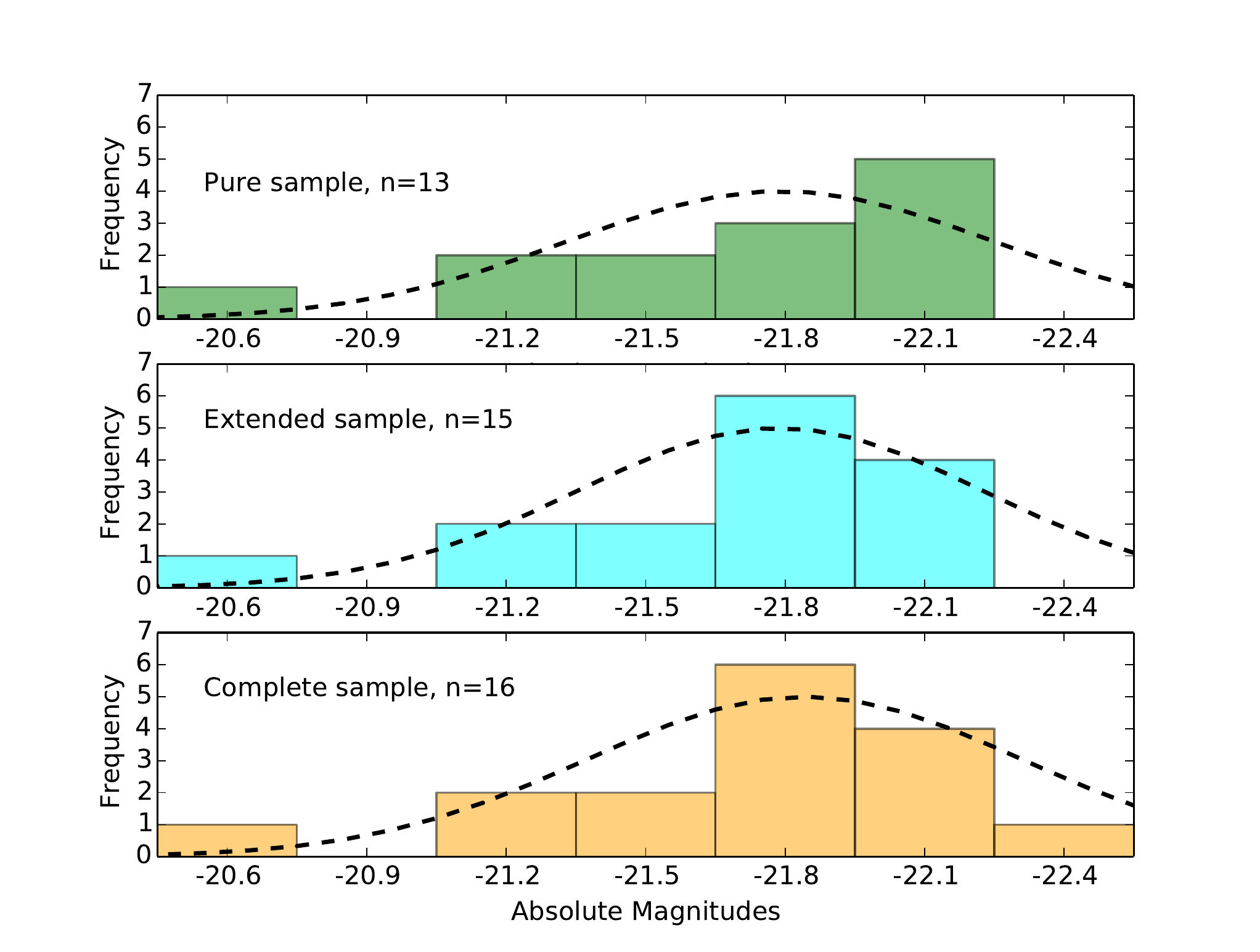}
\caption{Histograms of the raw (uncorrected) absolute peak magnitudes, $M(400)$,  for the three samples 
defined in Section\,\ref{sec:sample}. The dashed lines are least squares, best fit Gaussians. 
Assuming confidence limits for small number events from \cite{1986ApJ...303..336G}, the Gaussian fits all comfortably fall within the 1$\sigma$ errors of the frequencies. } 
\label{fig:histo}
\end{figure}

In Figure~\ref{fig:histo} the histograms of the absolute peak magnitudes of our samples are shown.
The numbers in each bin are understandably small, but the distribution is consistent with being  
approximately gaussian. Considering the statistical errors on the frequency in each bin
\citep[see small number confidence limits in][]{1986ApJ...303..336G}
we find no evidence for a non-Gaussian distribution of  peak magnitudes.

\section{The peak magnitude - decline rate relation}
\label{sec:rel}

The now famous Phillips relation for type Ia SNe links the rate of decline of a $B-$band SN Ia light-curve
to the absolute $B$ magnitude at peak \citep{ph93}.  The original paper by Phillips brought the intrinsic scatter in 
SNe Ia peak magnitudes down from the raw $\pm0.8$ in $B$ to $\pm0.36$, and Prieto et al.'s larger
sample produced a scatter of $\pm0.2$.
To determine if SLSN are standardizable in a similar way
and if the raw scatter {\bf of $\pm0.46$ mag}  can be improved upon, 
we correlated the absolute peak magnitude, $M(400)$, with the amount that the light curve fades from the maximum at three defined periods following peak light.

\begin{figure*}
\includegraphics[width=18.3cm]{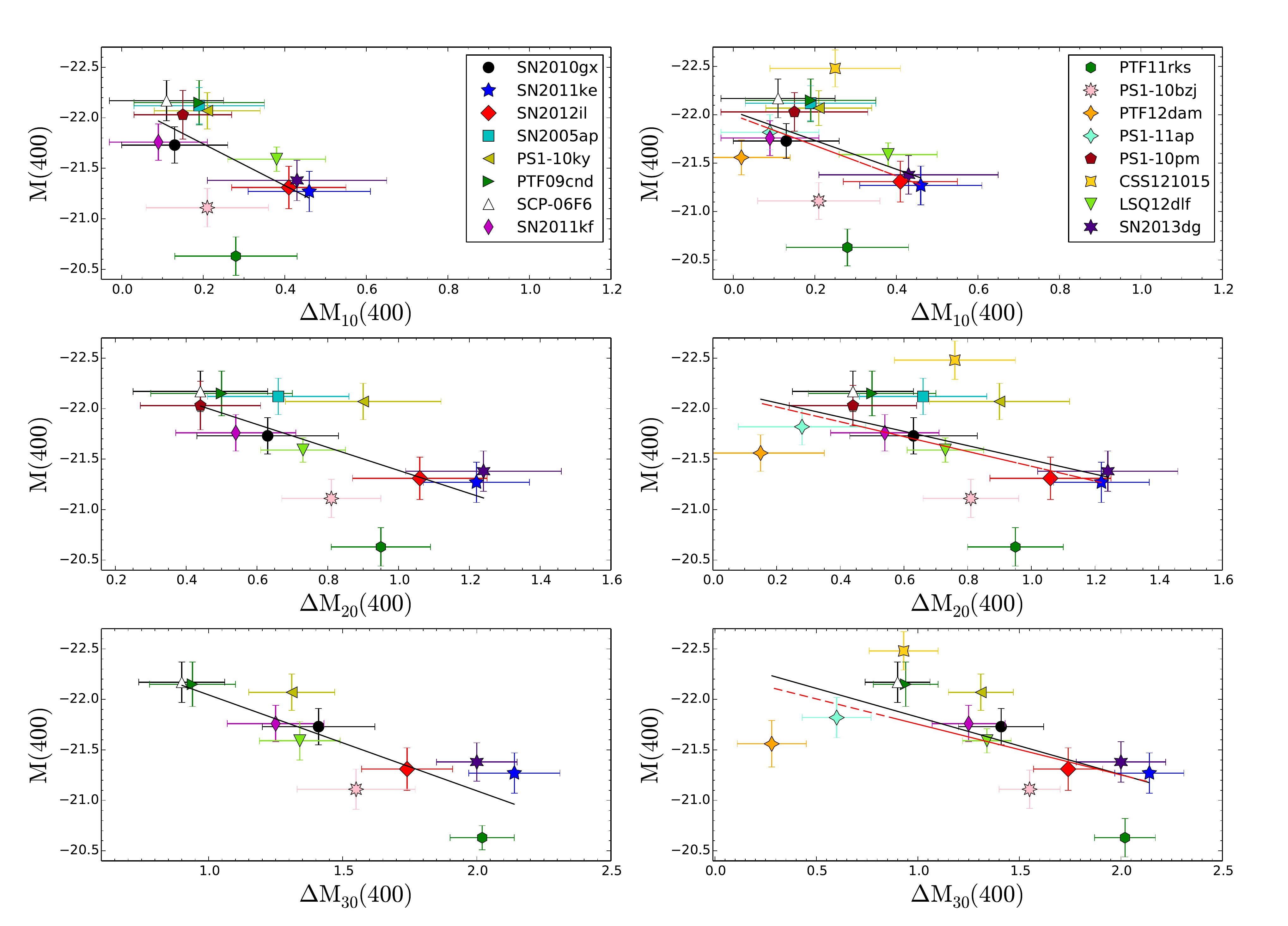}
\caption{The peak magnitude - decline rate relation. The absolute peak magnitude in the 400nm band, $M(400)$, is plotted versus $\Delta$M$_{10}$(400), $\Delta$M$_{20}$(400) and $\Delta$M$_{30}$(400). The latter three values are measures of the light-curve decline (in magnitudes) during the first 10, 20 and 30 days after maximum.  The left-hand column of plots shows only the 2005ap-like events (pure sample) and the 
right-hand column shows all SLSN. The dashed-red line is the fit  of the extended sample, while the black of the complete sample (see Section~\ref{sec:sample} for the sample definitions). 
The SN magnitudes and the corresponding $\Delta$M$_{\rm day}$(400) are reported in Table~\ref{table:data}, while the parameters of the linear regression fits (black line) are listed in Table~\ref{table:fit}.}
\label{fig:corr}
\end{figure*}

\begin{figure}
\includegraphics[width=\columnwidth]{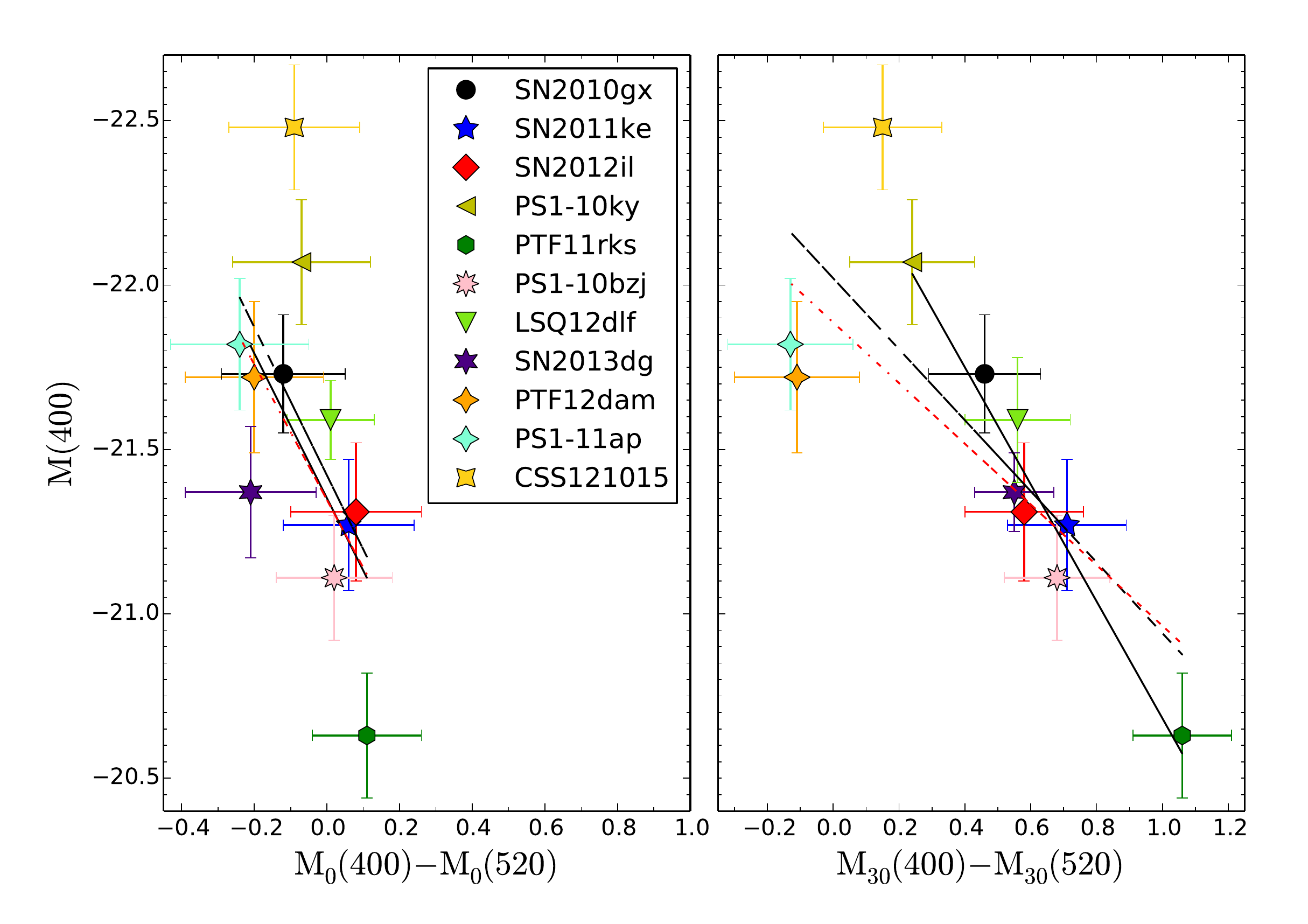}
\caption{The peak magnitude - color evolution relation. This plot illustrates the apparent dependency of peak magnitude on the color and color evolution rate. The left panel plots the $M(400)$ peak magnitude versus the color at peak brightness. The right panel plots the same
$M(400)$ at peak versus the color at +30 days (rest frame). The black line fit refers only to the pure sample (2005ap-like objects), the red dot-dashed line fits the extended sample (which includes the 2007bi-like objects) while the black dashed line fits the complete sample. } 
\label{fig:colevol}
\end{figure}

\begin{deluxetable*}{ccccccc}
\tablewidth{0pt}
\tablecaption{Fit parameters and statistical results of 
our pure, extended and complete sample\label{table:fit}}
\tablehead{\colhead{Days} & \colhead{N (objects)}& \colhead{a} & \colhead{b} & \colhead{$\sigma$ (mag)} & \colhead{Spearman}& \colhead{Pearson}}
\startdata
\cutinhead{$\Delta$M$_{\rm day}(400)$ 
pure sample}
10 &	13 &	2.06 (1.17)&	-22.15 (0.29) &	0.38 &	0.65 &	0.56\\
20 &	13 &	1.19 (0.47)&	-22.53 (0.37) &	0.34 &	0.73 &	0.66\\
30 &	11 &	0.95 (0.23)&	-22.99 (0.35) &	0.25 &	0.90 &	0.84\\
\cutinhead{$\Delta M(400-520)$ 
pure sample}
30 &	8 &	2.02 (0.42)&	-22.64 (0.26) &	0.19 &	0.71 &	0.88\\
\cutinhead{$\Delta$M$_{\rm day}(400)$ 
extended sample}
10 &	15 &	1.57 (0.99) &	-22.00 (0.22)&	0.38 &	0.51 &	0.49\\
20 &	15 &	0.73 (0.37)&	-22.16 (0.26)&	0.36 &	0.58 &	0.54\\
30 &	13 &	0.50 (0.20)&	-22.26 (0.27)&	0.33 &	0.75 &	0.64\\
\cutinhead{$\Delta M(400-520)$ 
extended sample}
30 &	10 &	1.09 (0.45) &	-22.00 (0.23) &	0.26 &	0.70 &	0.74\\
\cutinhead{$\Delta$M$_{\rm day}(400)$ 
complete sample}
10 &	16 &	1.49 (1.10) &	-22.03 (0.25) &	0.42  &	0.42 &	0.42\\
20 &	16 &	0.71 (0.42)&	-22.20 (0.30) &	0.41  &	0.51 &	0.47\\
30 &	14 &	0.57 (0.22)&	-22.39 (0.28)&	0.36  &	0.74 &	0.64\\
\cutinhead{$\Delta M(400-520)$ 
complete sample}
30 &	11&	1.34 (0.42) &	-22.19 (0.20) &	0.31 &	0.71 &	0.75
\enddata
\tablecomments{Least squares fits for an unweighted linear fit of the form $M_{\rm max}(400)~=~a\Delta M_{\rm day} (400)+b$ with uncertainties in parentheses. The $\sigma$ is the standard deviation of this fit. The last column gives the Spearman rank-order correlation coefficient and the Pearson correlation coefficient $r$. As the errors on each value of $M(400)$ are fairly similar, a weighted calculation is not significantly different. 
For completeness, we have reported the fit values obtained excluding CSS121015. Although SN2005ap and PS1-10pm are left out of the 30 day decline calculations as they do not have enough data.}
\end{deluxetable*}

We chose the timescales of 10, 20 and 30 days post maximum and refer to the difference between peak magnitude at these epochs as $\Delta M_{10}(400)$, $\Delta M_{20}(400)$ and $\Delta M_{30}(400)$ respectively. It is not possible to meaningfully test the extension of  the time baseline beyond  30 days as the sample size would reduce considerably due to lack of data at these epochs. 
The magnitudes of each SN at these epochs were estimated from the polynomial fits to the light-curve points
as described in Section~\ref{sec:int} with the only exception of the peak magnitude of CSS121015 (cfr.~Appendix\,\ref{app:fits}).  In Fig.~\ref{fig:corr} we plot the relation between $M(400)$ and the decline rates at 10, 20 and 30 days. 
The plots indicate that the peak magnitude of SLSN Ic
does appear to be correlated with decline rate in the 
same sense as for the Phillips relation for SNe Ia. The brighter the peak magnitude (in the rest frame 400nm band) the more slowly they fade. The question then becomes : is this quantitatively useful to reduce the intrinsic scatter in the 
uncorrected peak magnitudes below the {\bf raw scatter of $\pm0.46$ mag~?}

\begin{figure*}
\center
\includegraphics[width=18cm]{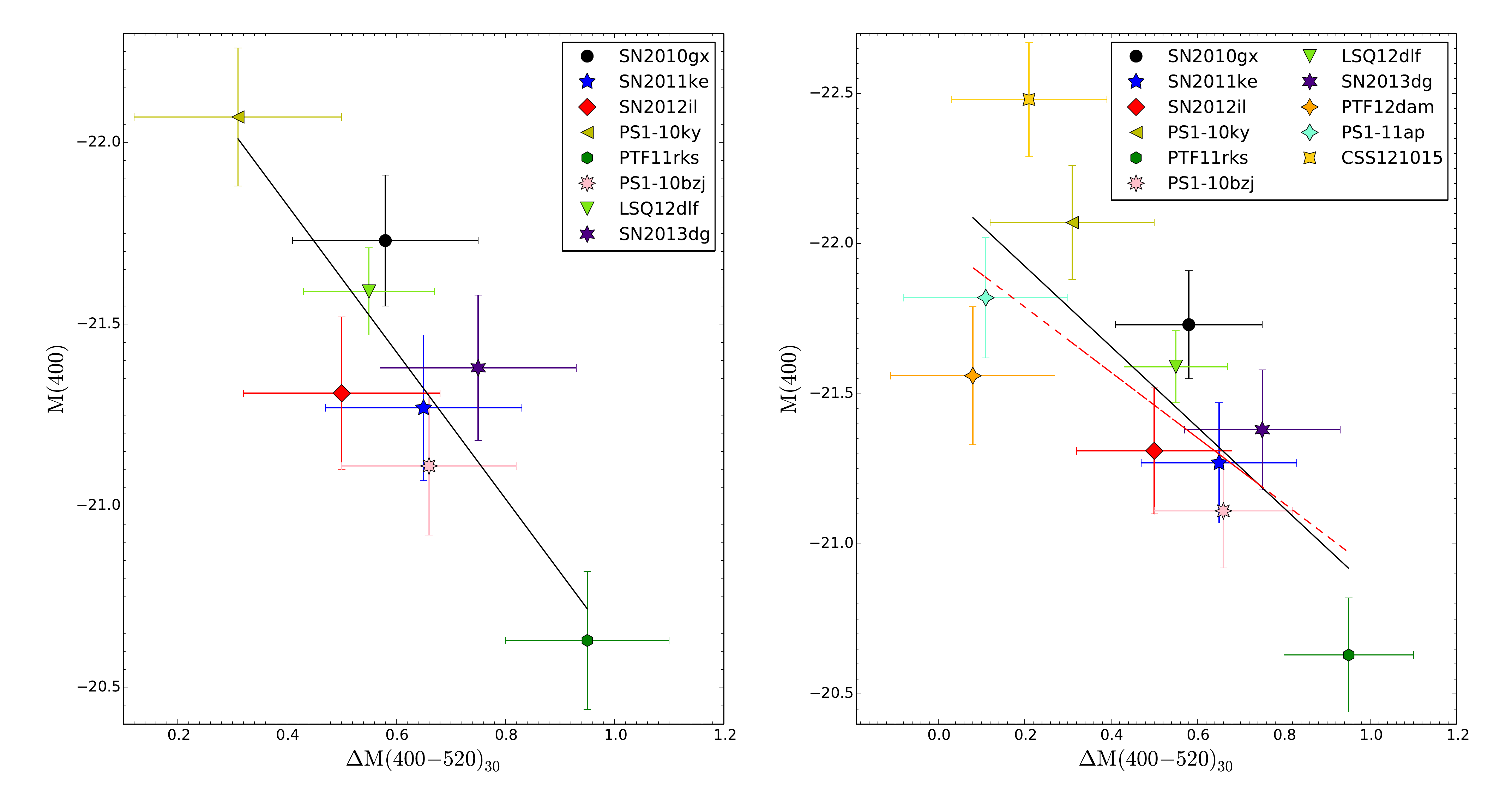}
\caption{The peak magnitude - color evolution relation. 
Absolute peak magnitudes in the 400nm band, $M(400)$, versus the color  change from peak to +30 days  (rest frame). We define $\Delta M(400-520)_{30} = (M_{0}(400) - M_{0}(520)) - (M_{30}(400) - M_{30}(520))$ The bands at 400nm and 520nm are defined in Section\,\ref{ss:kc}, and the values for $\Delta M(400-520)_{30}$ values are reported in Table~\ref{table:data}. {\em Left:} a plot of eight of the pure sample which 
have the necessary bands covered in the observer frame to calculate this color term. 
{\em Right:} eleven objects of the complete sample which have the necessary data. The black lines are least squares straight line fits to all data in each plot. The 
red dotted line leaves out CSS121015. The parameters of these line fits are 
are listed in Table~\ref{table:fit}.} 
\label{fig:colcorr}
\end{figure*}

The least square fits
to the data points are shown and their parameters are given in
Table~\ref{table:fit}. These suggest that the scatter 
can indeed be improved upon to between 
{\bf 0.25 and 0.33 depending} on the epoch at which the decline rate is applied and which object sample to include. 
The most promising relation that we find is that  
{\bf for $\Delta  M_{30}(400)$, which reduces the scatter from $\pm0.46$ mag to $\pm0.25$ mag using} the pure sample of SLSN Ic.
The correlation at 20 days after peak, $\Delta M_{20}(400)$, is {\bf similar to that at 30 days} . Although the rms scatter is formally larger, the difference is not particularly significant (0.03 to 0.09). 
{\bf Indeed, the} Spearman rank-order correlation coefficient for the $\Delta {\rm M}_{30}(400)$ rate is {\bf 0.90 while the Pearson's $r$ test gives a result of 0.84} 
and are also reported in full for each fit in Table~\ref{table:fit}. 

The Pearson correlation coefficient for {the $\Delta  M_{30}(400)$ is  higher than the $\Delta  M_{20}(400)$, }
and is usually the most commonly employed statistic. It measures the strength of the linear relationship between normally distributed variables. When the variables are not normally scattered, or the relationship between the variables is not linear, it may be more appropriate to use the Spearman rank correlation method.
Indeed, the Spearman rank correlation method makes no assumptions about the distribution of the data. 

For the extended sample
the scatter increases {\bf to 
$\pm0.33$ mag ($\pm0.36$ mag at 20d)} and the Spearman correlation coefficient and Pearson's $r$ are significantly reduced. 
This simple statistical result and visual inspection of the correlation plots in Fig.~\ref{fig:corr} would 
suggest that the 2007bi-like, slowly fading SLSN differ from the rest of the SLSN Ic. 
If they are included in the total sample, then it may result in distribution with an asymmetric tail and 
non-negligible  kurtosis. In this case we  should be careful in assuming a normal distribution for the 
parameters of all SLSN. It is not yet clear if PTF12dam, PS1-11ap and SN007bi are physically different explosions
to the rest of the SLSN Ic, nor is it clear if there is a continuum between the fast and slowly declining events.
The next slowest SLSN Ic are PS1-10pm and SCP-06F6, which do sit comfortably on all the straight line fits. 

If we would consider the complete sample 
then the scatter would {\bf increase to $\pm0.36$ mag (or $\pm0.41$ mag at 20d), } with a marginal decrease in the statistical correlation coefficients.  We 
would highlight that we are presenting the results grouped in these three ways simply due to the fact that the 
observational characteristics of the three groups are quite easily distinguished in this low-moderate 
redshift sample. However, if one were working at high redshift then the subtleties of the different 
samples and observational classes may be difficult to distinguish.

In Section\,\ref{ss:kc}  we described how the wavelength window for the synthetic 400nm photometric band was chosen. The
 spectra of all objects in this region are continuum dominated and relatively similar in slope and spectral features. Nevertheless there are  some absorption features  present and the O~{\sc ii} lines are typically the strongest features at $\lambda\geq4200$ \AA\/. These features are on the red edge of our window and show noticeable absorption before maximum light \citep{pa10,qu11,ch11}. 
At peak brightness, the O~{\sc ii} absorption at $\sim$4000 \AA\/ is relatively shallow and remains  weak in the following 20 days \citep{pa10,in13}. At 30 days past maximum, absorption due to 
Mg~{\sc ii} and Fe~{\sc ii} become noticeably stronger than the O~{\sc ii} lines.  
In our spectroscopic analysis we found two objects, namely PTF11rks and PS1-10bzj, that seem to show noticeable differences in their spectroscopic evolution with respect to the bulk of SLSN Ic. 
The  Mg~{\sc ii} and Fe~{\sc ii} lines between 3000~\AA\/ and 6000~\AA\/, and Si~{\sc ii} $\lambda$6355 are stronger 
in these two objects  than the rest of the sample.  In the first two weeks after peak, PTF11rks shows absorption line equivalent widths which are a factor of 4 greater than the bulk of the sample (EW$_{\rm PTF11rks}=4\pm1.5$~EW$_{\rm SLSN~Ic}$). The strengths of these features  
 are noticeable in Fig.\ref{fig:comp}.  The PS1-10bzj spectra after maximum light also show a qualitatively similar line intensity evolution,  although not as pronounced as PTF11rks. 
We note that the lower S/N of the spectra of PS1-10bzj prevent 
as detailed and  quantitative measurements as were possible for PTF11rks, but both appear to evolve on faster timescales than the rest of the SLSN Ic. For completeness, 
we tested the $\Delta M_{30}(400)$ relation {\em excluding} PTF11rks and PS1-10bzj from the pure sample. We found that the  scatter was substantially reduced {\bf to $\pm0.14$ and} we retrieved a Spearman rank-order correlation coefficient of {\bf 0.95 while the Pearson's $r$ test is 0.91}. The statistical results are slightly better than those obtained with  all objects of the pure sample. This 
might suggest that these two SNe appear to have spectroscopic and photometric differences 
compared to the bulk of the population. However,  identifying these potential outliers in the distribution is  too subtle to be achieved without a well-sampled time series of high signal-to-noise optical and near-UV spectra. Such 
identification would be difficult for high redshift objects. 

\section{The peak magnitude - color evolution relation}
\label{sec:colmag}

In Section\,\ref{ss:kc} we defined two synthetic photometric bands centred on 400nm and 520nm as regions of the rest frame spectra which were mostly devoid of strong SN absorption features and were, on the whole, continuum dominated.  The rest frame color evolution of the objects, based on these two photometric indices, is plotted in Fig.\,\ref{fig:colfit}.
This immediately illustrates a large diversity in color evolution which is visible at peak and increases substantially during the first 30 days. This is further illustrated in Fig.\,\ref{fig:colevol}, where we plot the absolute peak magnitude $M(400)$ against the color term $M(400) - M(520)$ at peak and after 30 days. 
The peak absolute magnitude $M(400)$ appears to be color dependent, in the sense that objects
which are fainter at peak are redder. If it were a real correlation, one would immediately 
consider internal host galaxy extinction as a possible explanation and we discuss this at the
end of this section.

A much more  striking trend  is that the fainter objects tend to become redder faster. In other words, the rate of color change appears to be correlated with peak magnitude. This strong dependency is displayed in Figure~\ref{fig:colcorr}, where we show that the peak absolute magnitude is correlated quite tightly with the {\em rate} of color evolution. In the pure sample, we are reduced to only eight objects due the requirement for data covering the rest frame 400nm and 520nm bands at both peak and at +30 days. This could potentially bias the result toward a positive result and artificially reduce the scatter around the straight line fit.  However the rms scatter around the linear least squares fit is reduced 
{\bf to  $0.19$ mag} for the eight objects of the pure sample. 
The two statistical correlation tests result in {\bf 0.71 for the Spearman coefficient and 0.88 for the $r$ Pearson's coefficient. }
For the ten objects of the extended sample the residual rms {\bf is still quite low at 
$0.26$ mag. It increases to $0.31$ mag for the complete sample}, when CSS121015 is included. 
From Figs. \ref{fig:colevol} and \ref{fig:colcorr} it is clear that PTF12dam and PS1-11ap are systematically bluer than the rest of the SLSN Ic both at peak and at +30 days, and CSS121015 also sits significantly above the best linear fit. 

For completeness, we also checked the correlation over the 20 day evolution timescale and found a poorer correlation. {\bf For 
$\Delta M(400-520)_{20}$ we found an rms of 0.30 mag (0.29 for the extended sample) and lower statistical coefficients than for the $\Delta M(400-520)_{30}$ (0.67 for the Spearman coefficient and 0.67 for the $r$ PearsonÕs coefficient).} Hence the
30 day color rate evolution seems significantly better. 
Although the numbers are low, this peak magnitude - color evolution relation appears to be quite promising as a method of standardizing the magnitudes of SLSN. 
The rms scatter is surprisingly low, 
even with the  inclusion of the most outlying three objects in the complete sample. 
The very tight correlation for the eight SLSN Ic of the pure sample makes it imperative that this relationship is tested with a larger sample.

The  major uncertainty is extinction within the host galaxies. We have made no correction for internal 
dust extinction and it is difficult to quantitatively measure reddening from either the colors,  spectrum slope or interstellar medium lines in this small sample (indeed we do not detect ISM  Na\,{\sc i} lines 
from the ISM in the host galaxies for 
any SLSN).  One might be concerned that the left panel in Fig.\,\ref{fig:colcorr}
was  a reddening effect if we simply observed that the fainter SLSN were redder at peak. 
An argument against
it being extinction is that both PTF11rks and PS1-10bzj have qualitatively different 
evolution in their spectral features (deeper Mg\,{\sc ii}, Fe\,{\sc ii} and Si\,{\sc ii} absorption)
than the rest of the pure sample. This suggests cooler  photospheric temperatures rather
than an extinction effect  (see discussion in Section\,\ref{sec:rel}). 
The brightest object (CSS121015) is also spectroscopically different in that H$\alpha$ was
detected and it may well be physically different to the others.  Constant extinction values might explain the connection between color and luminosity, but they cannot easily account for the observed color evolution in the light curves.
The distinctive time evolution of the $M(400)-M(520)$ color could, in principle, be due to time-variable internal dust extinction. However, to our knowledge, such time variability has not been observed before toward any supernova. Time variable interstellar and circumstellar medium absorption lines have been observed toward SNe Ia \citep{2007Sci...317..924P}, but these have not been linked with variable extinction toward the SN photosphere.
In the future, a more sophisticated approach would be to combine fits for decline and color evolution into one standardization procedure as has been successfully done for SNe Ia  e.g.  methods like SALT2 \citep{guy07}. Or to use Bayesian methods to simultaneously model  dust and color effects as  in   \cite{man09,man11}. 
Additionally we plan to look for extinction signatures by searching for objects with identical
photospheric spectra but which have different colors and continuum slopes. This would 
ideally include near infra-red photometric fluxes to anchor the SED fits securely. Both of these
approaches require a more extensive data set than we have presented here, which should be 
possible given the ongoing low redshift surveys such as 
 PESSTO and La Silla QUEST \citep[][Smartt et al. in prep.]{2013PASP..125..683B}, CRTS \citep{dr09} and iPTF \citep{2013ApJ...775L...7C}.

\subsection{A control test : application to Type Ib/c}

\begin{figure}
\vspace{0pt}
\includegraphics[width=\columnwidth]{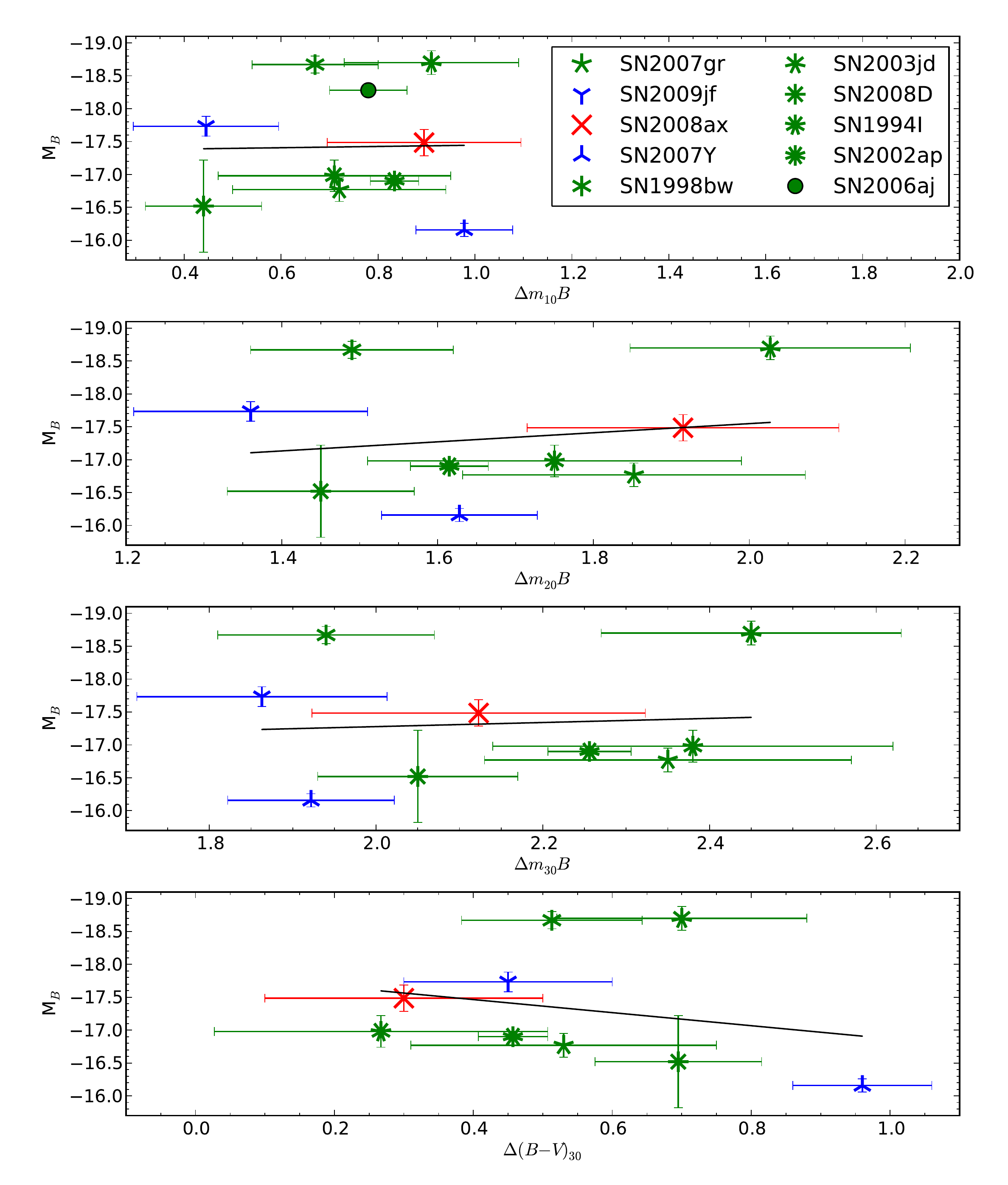}
\caption{ The  peak magnitude -  decline relation and the  peak magnitude  - color evolution relation for ten stripped envelope CCSN. Absolute magnitudes in the $B-$band are plotted versus
$\Delta m_{10}(B)$, $\Delta m_{20}(B)$ and $\Delta m_{30}(B)$ 
analogous to the SLSN sample in Fig.\,\ref{fig:corr}. The  $\Delta (B-V)_{30}$,  measures the color evolution in $B-V$ over 30 days.
 The data sources are as follows  :  the type Ic SN 2007gr \citep{va08b,hu09}, 1998bw \citep{ga98,mck99,so00,pat01}, 2003jd \citep{va08a}, 2008D \citep{ma08,sod08,mo09}, 1994I \citep{ri96}, 2002ap \citep{pa03,fo03,yo03,to06} and 2006aj \citep{ca06,co06,mi06,pi06,so06}; the type Ib SN 2007Y \citep{st09}, 2009jf \citep{va11} and the type IIb SN 2008ax \citep{pa08}.} 
\label{fig:corrIcIb}
\end{figure}

As a sanity check we investigated if the relations we appear to have discovered for SLSN Ic also manifest themselves in the normal type Ib/c SN population and hence if we are recovering characteristics which are  commonly displayed in more normal hydrogen poor stellar explosions. 
We performed the same analysis with a sample of well observed stripped-envelope SN (type Ic, Ib, IIb), with the only difference from the SLSN Ic analysis being the choice of the reference bands. We chose the Johnson $B-$band with central peak wavelength at 4448 \AA\/ and  Johnson $V-$band 5505 \AA, due to the excellent sampling of the low-redshift observed SN Ib/c in this standard wavebands (which were not in need of any $K$-correction).
As shown in Figure~\ref{fig:corrIcIb} there is no relation between the peak magnitudes and the decline rates
at 10, 20 and 30 days or with the color evolution. As expected, we find a large scatter of peak magnitudes and no obvious correlations which mimic the findings for the SLSN. The mean and raw scatter in the $M_{B}$ values for type Ibc SN are $-17.42\pm0.90$, and this is not improved when any linear fits are applied. 

\begin{figure*}
\includegraphics[width=18cm]{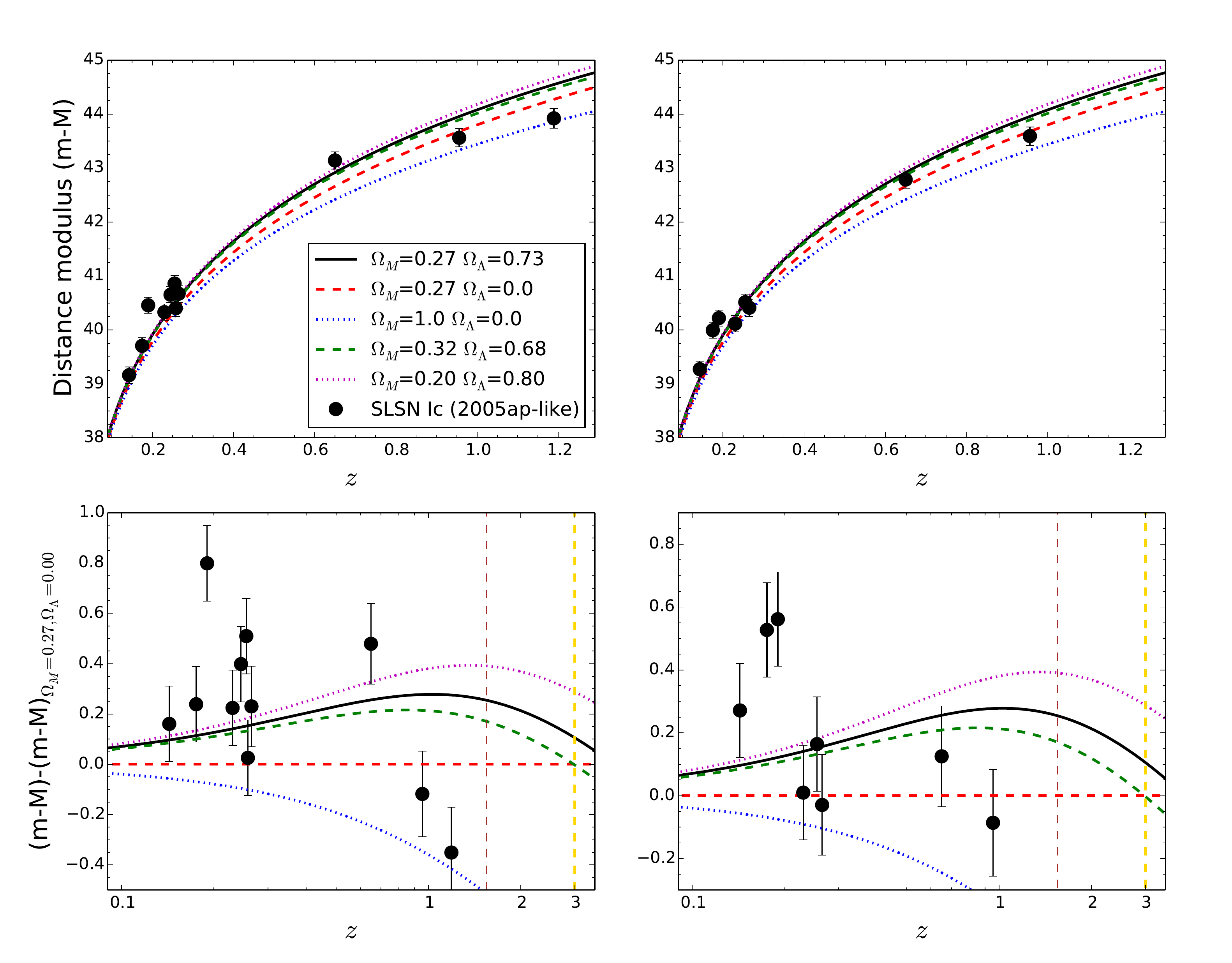}
\caption{Hubble diagrams for SLSN Ic. Different cosmologies are plotted in comparison with that chosen (solid black line) to measure the distance moduli of the sample presented. 
{\em Left panels:} The distance moduli measured using the  $\Delta M_{30}$ decline relation 
{\em Right panels:}  The distance moduli  measured using the peak magnitude  - color decline relation.  The  lower panels show the residuals of the distances relative to a $\Omega_{\rm M} = 0.27$, $\Omega_{\rm \Lambda} = 0$ Universe. The brown, dashed vertical line represents the SNe Ia upper limit with the current generation of telescopes (ground and space), whereas the gold, dashed line represents an approximate  limit for discovery and monitoring of SLSN Ic with the current generation of telescopes.} 
\label{fig:hubble}
\end{figure*}

\section{Hubble Diagram and Residuals}\label{sec:h}
If we assume that SLSN Ic can be used as standardizable candles and then apply the $\Delta M_{30}$ decline  relation (calibrated with 11 objects) and the peak magnitude - color evolution relation (with the 8 objects reported in Sect.~\ref{sec:colmag}), then we can use these as a preliminary test to check if they fall in a reasonable region of the Hubble diagram, as shown in Figure~\ref{fig:hubble}. 
The small number of objects and lack of a low redshift anchor prevents any detailed discrimination between cosmology models 
but consistency checks of the relative distances for a cosmology with $\Omega_{\rm \Lambda} = 0$ is possible. 

We evaluated the distance moduli with the simple formalism $\mu=m-M+\alpha\;\Delta M$, where $\Delta M$ refers to the standardization relation used and $\alpha$ is a free parameter.  This parameter $\alpha$ is allowed to vary and to minimise the $\chi^2$ between the fit to the distance moduli and those for different cosmologies. Because of our statistically limited dataset, we can only determine if the
data are consistent with a  $H_0=72$, $\Omega_{\rm M} = 0.27$, $\Omega_{\rm \Lambda} = 0.73$ universe. We found 
a satisfactory fit, and the lowest $\chi^2$ for this concordant cosmology model. If we attempt a fit with $\Omega_{\rm \Lambda} = 0$, 
$\Omega_{\rm M} = 1$,  then we get a $\chi^2$ which is a factor of 10 worse, and an unsatisfactory Hubble diagram fit. 
This simply illustrates that the relative distances to these SLSN are not compatible with a cosmological model with 
$\Omega_{\rm \Lambda} = 0$, but we do not have enough statistics to test it any further. 

The results of the best fit are displayed in Figure~\ref{fig:hubble}. It also indicates that distance indicators in the range of $0.6 \lesssim z \lesssim 3$, where SLSN Ic are already discoverable with the current generation of telescopes, may give us some  leverage on 
$\Omega_{\rm M}$ since the difference between $\Omega_{\rm M}=0.20$ and 0.32 is  $\sim0.3$ mag which would be well within the reach of a sample of $10-20$ SLSN Ic in the redshift range $z=2-3$. 
SLSN will probably not achieve a competitive precision to {\bf rival  $\Omega_{\Lambda}$ measurements} 
from  either Baryon Acoustic Oscillations \citep{eis05,per10} or Cosmic Microwave Background 
experiments \citep{dun09,kom11} as already  noted by \citet{su12}. However since SLSN are the only current alternative to SN Ia for estimating radial distances at high redshift, it is worth exploring if they 
can constrain the time-varying nature of $w(z)$, the dark energy equation-of-state parameter.
The study of SLSN at high$-z$ is also a promising avenue to explore 
the nature of the explosions and sources of luminosity and to investigate if there are changes in their
properties with epoch and environment (see Appendix~\ref{app:prop}).

In the future, we plan to tie down the calibrations for these potential standardizable candles in the lower redshift regime of 
$0.1<z<0.3$ simply by increasing the number of objects discovered and monitored in detail. Their cosmological usefulness may stem primarily from their ability to be discovered and studied in the
redshift regime beyond that for SNe Ia, which is 
currently around $z=1.55$ \citep{2012ApJ...746....5R}. 


\begin{figure*}
\center
\includegraphics[width=17cm]{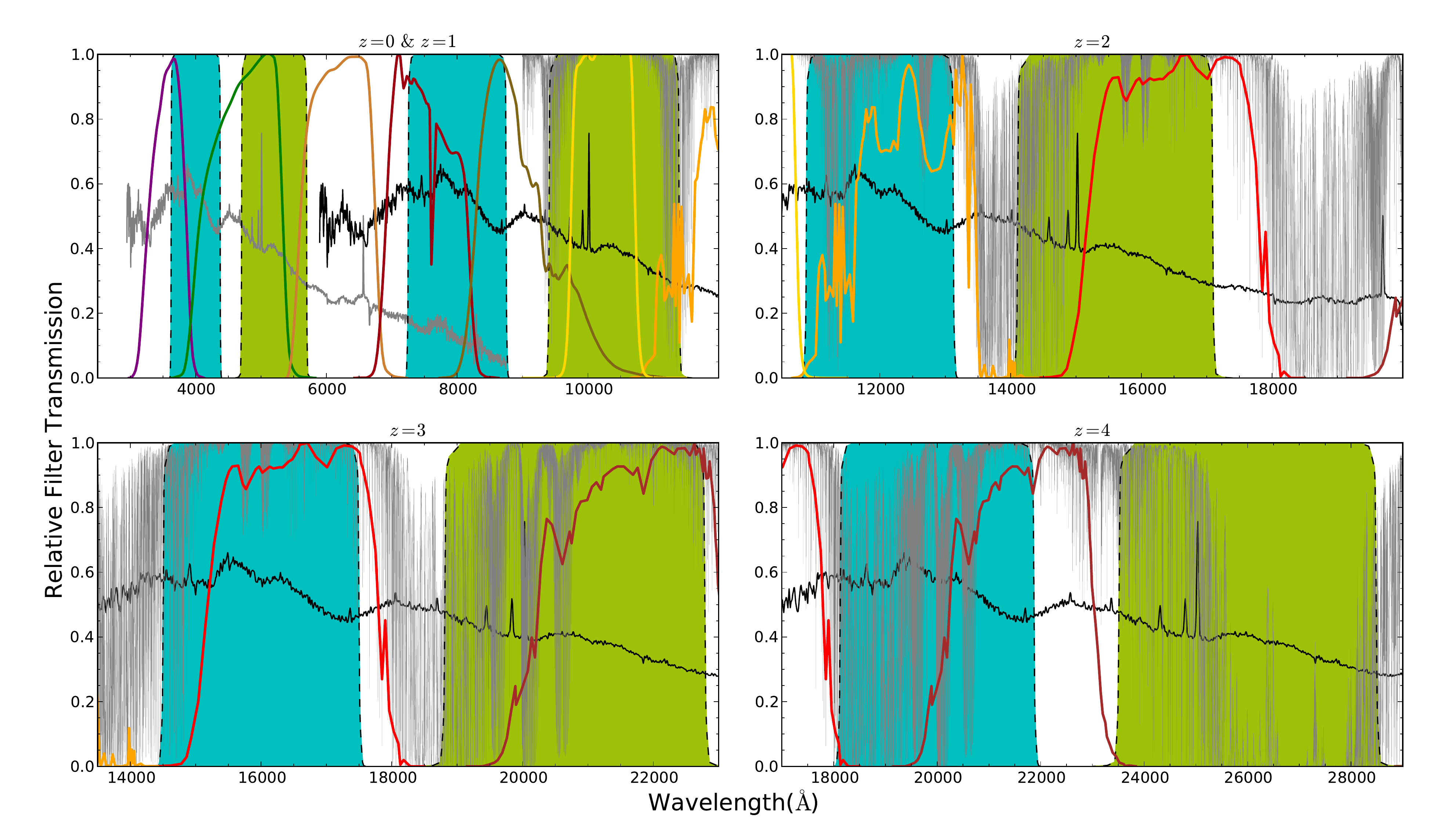}
\caption{
Synthetic passbands 400nm (cyan) and 520nm (green) plotted with SN2011ke spectrum at $\sim$10d scaled in flux (grey) at rest frame and at projected distances of $z=1$ (upper left corner), $z=2$ (upper right corner), $z=3$ (bottom left corner) and $z=4$ (bottom corner). Sloan $u$ (purple) $g$ (green), $r$ (bronze), $i$ (burgundy) and $z$ (oak brown) bandpasses are displayed along the near infrared $Y$ (gold) and the 2MASS $J$ (orange), $H$ (red) and $K$ (brown) bandpasses. Also plotted in grey is the atmospheric transmission in the NIR \citep[][courtesy of Gemini Observatory]{lo92}.} 
\label{fig:ftran}
\end{figure*}

\begin{figure}
\center
\includegraphics[width=\columnwidth]{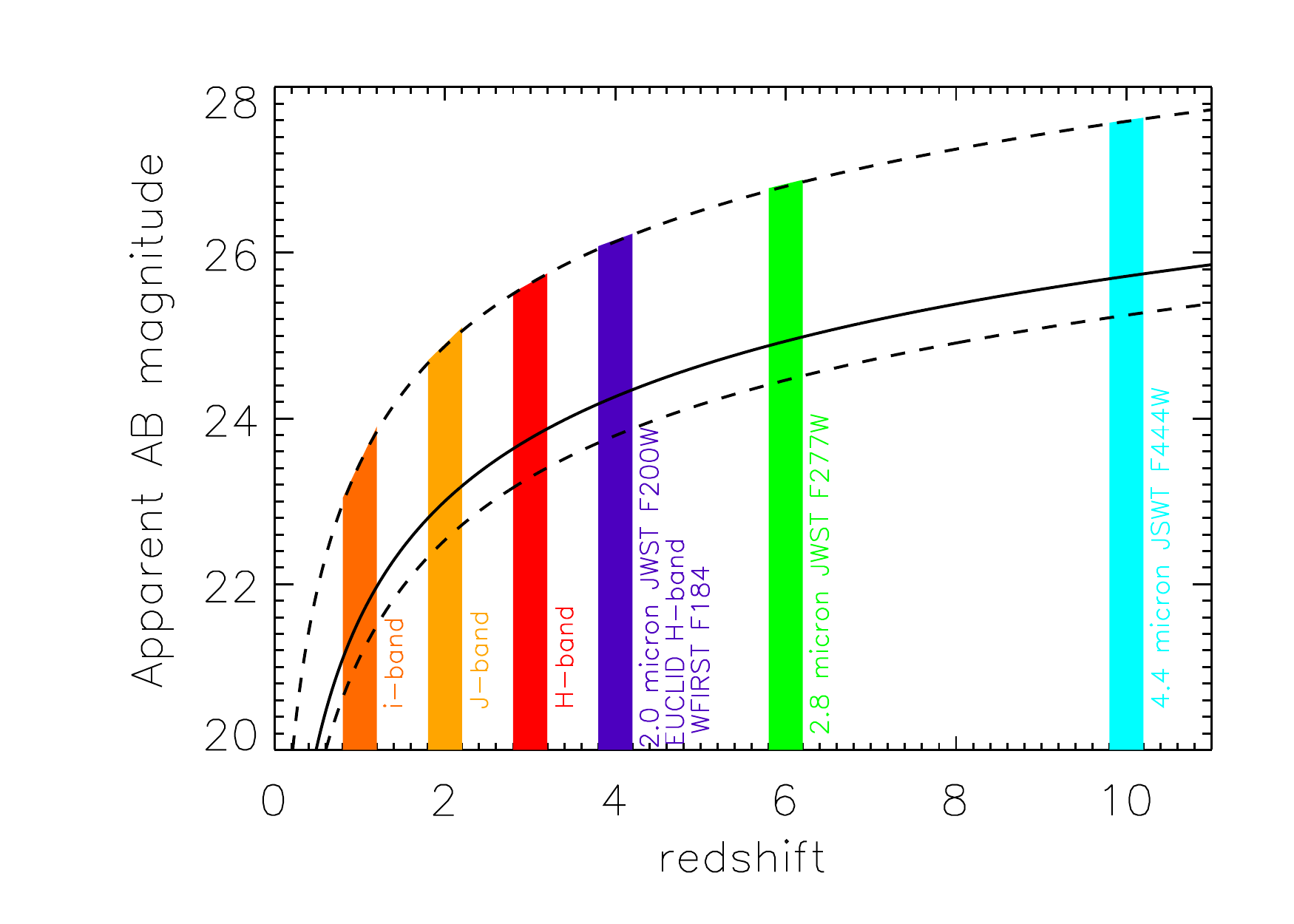}
\caption{At higher redshift the only possibility of observing the 400nm and 520nm bands is using future space based telescopes. This figure illustrates the predicted AB magnitude for SLSN Ic out to $z\simeq10$. The solid line is the peak magnitude {\bf of $M(400)=-21.70$ derived here. The lower dotted line is 1$\sigma$ scatter on the peak magnitudes and the upper dotted line illustrates the depth required to reach  +30 days after peak, to apply the $\Delta M_{30}(400)$ correction discussed here (addition
of the typical 1.4$^{m}$ decline }. No foreground extinction has been applied to the estimated magnitudes, since the Milky Way foreground will be minimal at these wavelengths.} 
\label{fig:highz} 
\end{figure}

\section{Future use of SLSN as high redshift distance probes}\label{sec:rate}

If our results are confirmed with a larger sample then these SLSN show potential as new cosmological probes stretching well 
into the the epoch of deceleration. At $z\geq1$, the 
synthetic passbands at 400 and 520 nm move into the near-infrared. We primarily chose the central wavelength and width of these
rest frame bands due to the lack of strong features in the SLSN Ic spectra (see Section~\ref{ss:kc})  but we also considered the
location and width of the redshifted passbands in the observer frame at $1\lesssim z \lesssim 4$. 
In Figure~\ref{fig:ftran} we show the two bands redshifted at  $z=1,2,3,4$. The 400nm band covers a similar wavelength region to the Sloan $i$-band, the NIR $J$-band, $H$-band and partially the $K$ band, respectively.  Observations in these bands 
would mitigate the dependence of the cross-filter 
 $K$-correction on the spectral template used. Similarly, the 520nm band would be reasonably well 
covered at $z=1,2,3$ by the  standard filter bands $Y$, $H$ and $K$ , respectively. No standard ground-based filter band is found  to match the 520nm band at  $z=4$ due to atmospheric absorption from $\sim2.5\mu$m to $\sim3.5\mu$m. As an example, 
typical average magnitudes of SLSN in  the rest frame 400 nm band would correspond to $J\sim23.0$, $H\sim$23.8 and $K\sim$24.3 mag at $z=2,3,4$ respectively. The exposure times with the VLT and HAWK-I are between 2-4hrs to reach
a signal-to-noise of 10 for these magnitudes.  While the Wide Field Camera 3 (WFC3) on the Hubble Space Telescope 
can only get to $J$ and $H$ equivalent wavelengths (F110W and F160W), one can reach improved photometric precision 
(S/N$\simeq$50) in manageable exposure times of between 300s to 3000s on source. 
While these are feasible as dedicated follow-up projects, neither HAWK-I nor WFC3 are appropriate survey instruments to find them in large numbers due to their limited field size (tens of square arc minutes). A more appropriate search engine is a planned survey such as ``SUDSS" (Survey Using Decam for Super-luminous Supernovae) using the Dark Energy Camera on the CTIO Blanco 4m telescope. This project has a goal of discovering 200 SLSN out to $z\simeq4$ over 3 years by imaging approximately 24 square degrees (in $griz$) to $i\simeq25$ every 14 days.  The planned 
Subaru/Hyper Suprime-Cam deep survey will also target these high redshift objects, with 
encouraging numbers estimated by  \cite{2012MNRAS.422.2675T}. 
The NIR imaging follow-up would be required 
 to reach the rest frame 400nm and 520nm bands and exploit our proposed standardizing of the peak magnitudes. 
Identification of the redshifts would not require spectra in the NIR, but could be done in the optical observer frame  (rest frame
UV wavelengths between 1200-2500\AA) where the SLSN are predicted to peak around $i=23.5$ at $z\simeq3$. 

Despite their enormous luminosity and potential for standardizing the peak magnitudes, two practical difficulties 
remain. The first is that we have  only eight to thirteen SLSN (depending on sample selection) having sufficient multi-color data at the epochs and filters required, thus increasing the sample is mandatory. The second is their low volumetric rate of cosmic production. 
A rough estimate of their rate  in the local universe was provided by \citet{qu13b}, who reported a rate of $32^{+77}_{-26}$ events h$_{71}^3$ Gpc$^{-3}$ year$^{-1}$ at a weighted redshift $\bar{z}=0.17$ although was based on only one detected object (SN2005ap) in the Texas Supernova Search (TSS). While the number is uncertain, the TSS and PTF \citep{qu11} do demonstrate that  the SLSN rate in the local universe is low compared to that of SN Ia (0.01\%). \cite{mc14b} has estimated a rate of 
$6^{+3.6}_{-2.4}\times10^{-5}$ and $1^{+2.3}_{-0.8}\times10^{-4}$  of the CCSN rate within $0.3\leq z\leq1.4$ 
However at redshifts $z\geq1.5$, \cite{co12} suggested that 
the rate could be significantly higher,  possibly as a consequence of decreasing metallicity and increasing cosmic star formation rate density. SLSN Ic may require low metallicity to be be produced
\citep[e.g.][]{ch13,lu13b} but the efficiency of producing a SLSN from progenitor systems is still unknown and the progenitor stellar mass range  is undetermined. Future surveys exploring the low, medium and high redshift distances are necessary to strengthen, or falsify, the findings here reported and hence make use of SLSN as high redshift distance probes.

Looking further ahead, detecting SLSN Ic  out to redshifts of $z\simeq10$ is quite plausible but requires space based surveys to sample the restframe 400nm and 520nm  bands proposed here. As an illustration of what would be required, we plot the apparent AB magnitudes of SLSN Ic 
between redshifts 1 and 10 in Fig\,\ref{fig:highz}. These are the redshifted $M(400)$ magnitudes in the AB system, from the conventional definition 

\begin{equation}
m_{\rm AB} = M(400) + 5\log(D_{\rm L}(z)/ 10 {\rm pc}) - 2.5\log(1+z)
\end{equation}

where $M_{\rm AB}$ is the apparent AB magnitude at the redshifted wavelength and 
$D_{\rm L}$ is the luminosity distance with our chosen cosmology. 
We label the wavelengths and filter systems required to cover the 
restframe 400nm band at specific wavelengths.  The magnitudes are well within the capabilities of 
the future EUCLID\footnote{http://www.euclid-ec.org} 
\citep{2011arXiv1110.3193L}
, WFIRST\footnote{http://wfirst.gsfc.nasa.gov/} and James Webb Space Telescope (specifically NIRCam\footnote{http://www.stsci.edu/jwst/instruments/nircam/}) 
missions, even out to 
redshifts of $z\simeq10$. However, the luminosity distances at these redshifts are exactly what we would like to probe using our standardisation for the low
redshift objects. 
A major uncertainty will be the rates and the numbers that these
space missions may discover in their general survey and custom designed survey modes
\citep[see][for  initial estimations of detection rates of high redshift SNe]{2012MNRAS.422.2675T,co12}.

\section{Conclusion}\label{sec:conc}

From a study of 16 well observed and well sampled SLSN Ic in the redshift range $0.1 < z < 1.2$ we have investigated their use as standard  and standardizable candles. We 
defined two synthetic photometric wavebands and applied quantitative $K-$corrections from observed spectra to 
compare their magnitudes in consistent rest frame wavebands. These two wavebands are 
centred on 400nm (with a width of 80nm) and 520nm (with a width of 100nm) and were chosen because the spectra are 
continuum dominated and relatively free from absorption features.
 The raw scatter of the uncorrected $M(400)$ absolute peak magnitudes is $\pm0.47$ mag which is immediately of interest for cosmological use, given that they have a mean peak magnitude of $M(400)=-21.70$ mag, around two magnitudes brighter than SNe Ia. 
We have proposed  that they can be standardized further by comparing their
decline rates at {\bf +30d after peak}, in the rest frame in a similar manner to the 
Phillips $\Delta m_{15}$ relation. 
It appears that the fainter the SN, the  faster the decline rate. {\bf The raw scatter can be  reduced 
to $0.25-0.33$\,mag using this $\Delta M_{30}(400)$ relation,} 
depending on which sample and calibration method is used. 

We have also uncovered an additional method to standardize the peak magnitudes, by comparing the rest frame color evolution of SLSN. 
Comparison of the rate of change of color index $M(400)-M(520)$ with the peak magnitude {\bf $M(400)$ shows a tight correlation with a very small rms scatter of
$0.19-0.26$\,mag.} The fainter SNe are redder at peak, and they evolve to 
redder colors faster than their bluer and brighter counterparts.
While this low rms is encouraging, we caution that it may be due to small numbers of objects since we could only use eight to ten of the SNe due to data limitations. An immediate objective is to increase the sample with enough observational data that this can be tested further. 

A difficulty remains in that SLSN Ic may have distinctly different observational (or physical) sub-classes. The bulk of the population are split into two broad classes of fast evolving objects (2005ap-like) and more slowly evolving objects
(2007bi-like). It is not immediately clear if these are distinct or there is a continuum
of properties bridging the gap between them. The residuals of the fits in our
standardizable candle methods are somewhat dependent on classification 
of these SLSN into these sub-classes. However the results are not critically
dependent on this phenomenological classification. A further complication is the
recent discovery of a SLSN which is spectroscopically similar to the SLSN Ic, but shows additional weak $H\alpha$ emission  (and hence has been classified as a SLSN type II). As it may be difficult to deselect this type of SN at high redshift from a 
pure sample of SLSN Ic, we included this event in our sample to check if it compromised the results. While it does appear brighter and bluer than the 
bulk of the SLSN Ic sample, its inclusion did not significantly increase the 
scatter in the {\bf $\Delta M_{30}(400)$ relation}. It did however increase the 
scatter in the peak magnitude - color evolution calibration {\bf to  $0.26$\,dex. }

The low volumetric rates of these SLSN means that they are not useful as primary distance 
indicators in the local Universe ($z<0.1$). Indeed, the calibration of their peak magnitudes
requires an adoption of a cosmology to determine their distance moduli at low to moderate 
redshift. Their usefulness may be in redshift regimes beyond those possible for SNe Ia. 
With current instrumentation and facilities, they have the potential to probe the Universe at $z=2-4$, in which their rest frame $M(400)$ and $M(520)$
calibration bands correspond to apparent AB magnitudes of between $23-24$  in typical near-infrared $JHK$ passbands. They are likely to be detectable out to redshifts of $z\simeq10$ with future
space based missions such as JWST, EUCLID and WFIRST. 

\acknowledgments
We thank the anonymous referee for a careful review which improved the clarity of the paper.
We wish to thank Mark Sullivan, Bob Nicholl, Andrea Pastorello, Enrico Cappellaro, Anders Jerkstrand, Morgan Fraser, Ting-Wan Chen, David Young, Stuart Sim, Rubina Kotak and Matt Nicholl for the helpful discussions. 
CI also thanks Robert Quimby, Ragnhild Lunnan and Stefano Benetti for providing the spectra of SN2005ap, PTF09cnd, PS1-10bzj and CSS121015. CI thanks Fulvio Melia and his team for the careful reading of the paper and their suggestions.
The research leading to these results has received funding from the
European Research Council under the European Union's Seventh Framework
Programme (FP7/2007-2013)/ERC Grant agreement n$^{\rm o}$ [291222] (PI
: S. J. Smartt).

\appendix

\section{Details of the polynomial fitting and interpolation of light-curves}\label{app:fits}

When fitting polynomials to the photometric lightcurves, we tried as far as possible to use the same time baseline: from $-10\pm5$ days before maximum (in the 400nm band) to $40\pm5$ days after. The fits displayed in Figures~\ref{fig:lcfit}~\&~\ref{fig:colfit} make use of this baseline and the magnitudes reported in Table~\ref{table:data} are those from interpolation at the specific epochs of peak, +10d, +20d and +30d, unless otherwise noted (below, we refer to these as the key epochs). 
Further details on each object are as follows. 

\begin{itemize}
\item SN2010gx: This is a very well sampled SN light-curve with only two small gaps around 20 days and 27 days past maximum.  We fitted 31 epochs from -8d pre-peak to 40 days after with a third order polynomial (rms = 0.06). 
For the key epochs we used magnitudes from interpolation, which were coincident with the measurements available specifically at epochs 0 and +10d.  A polynomial of order 3 was used to fit the $M(400)-M(520)$ color curve with an rms = 0.05 in the period -5d before peak to 40 days after.
\item SN2011kf: We lack specific data close to two out of four key epochs (+10d and +30d), but the light curve is extremely similar to that of SN 2010gx. The lightcurve is quite well sampled from +40d onwards \citep{in13}  and we also used these points to constraining the fit. We fitted the 6 epochs available until 45d post maximum with a second order polynomial. There is no noticeable difference between a second and third order function. We found an rms = 0.05 for the second order. 
If we would extend the data used out to 60 days, then the difference is in the rms fit is less than 2$\sigma$. The magnitudes reported in Table~\ref{table:data} are derived from the fit.
\item SN2011ke: This object has a well sampled $g-$band light-curve in the first 30d post maximum. Small gaps exist at +5d, +17d and +27d after peak. We have measurements specifically at all four key epochs and the magnitudes from interpolation are in agreement within the fit rms, hence in Table~\ref{table:data} we reported the actual magnitudes. An rms = 0.07 was found for a third order polynomial fit to the 19 epochs available for the baseline $-17$ to 31 days. If we extend the baseline up to 60 days post maximum and then include other 3 epochs, the rms for the same fit order of before increases to 0.08. We also fitted the $M(400)-M(520)$ color evolution with a third order polynomial  (rms = 0.07) from $-17$ days before peak to 31 days after. The rms of the fit does not change if we include the following point at +51d.
\item SN2012il: For this object, an observed $g-$band magnitude at peak is not specifically 
available but we do have an $r-$band measurement exactly at peak.  To retrieve a $g-$band peak magnitude 
we used our library of template spectra at peak epoch and the observed $r$ magnitude at peak to obtain the $K$-correction to the 400 nm rest frame absolute magnitude. To check the reliability of this method we also fitted the available photometry (from $r$ to $z$) around peak epoch with a blackbody and then estimated the $K$-correction factor from $r$ to 400 nm. The values retrieved with these two methods (template and multi-color photometry) were in agreement \citep[see][for further details on these methods and their application]{in13} within an error of $\sim0.11$ mag. The peak magnitude in 
Figure~\ref{fig:lcfit} is that from the spectral $K-$corrected method. For the $g-$band data available (+4d to +34d)
we used a third order polynomial fit, obtaining an rms = 0.04. The fit displayed in Figure~\ref{fig:lcfit} does not 
use the peak value, but there are measurements close to the key epochs of +10, +20 and +30d.
We used the same time baseline of the light curve for the $M(400)-M(520)$ color but we fitted it with a second order polynomial to avoid over fitting.
\item SN2005ap: We fitted 15 epochs  from $-5$ days from maximum to 26 days after with a third order polynomial which resulted in an rms = 0.09. The point at 19d past maximum was 4$\sigma$ from this fit and we removed it during the fitting 
process in accordance with  Chauvenet's Criterion.
\item PS1-10ky: This SN did not have a magnitude specifically at peak in the observed $i$-band (chosen for the 400nm rest frame cross-filter $K-$correction), however it did have  $r$-band measurements.  
We fitted 15 epochs of $M(400)$ (computed from observed $i$-band) between +6 and +36 days after maximum with a third order polynomial (rms = 0.07). The values listed in Table~\ref{table:data} refer to magnitudes from interpolation with this
fit.  To check the consistency of our fit we also specifically evaluated the peak magnitude with a $K$-correction from the observed $r-$band at peak to the 400nm rest frame band, together with a spectrophotometric measurement. 
As in the case of SN2012il these two methods gave back similar results. The magnitude  retrieved is plotted in Figure~\ref{fig:lcfit} and it matches the fit well. Hence we are confident that the absolute peak magnitude is reliable.  
While fitting the $M(400)-M(520)$ color evolution,  two points at $\sim$8 and $\sim$22 days past maximum were rejected by application of Chauvenet's criterion (being more than 3$\sigma$ from the fit). The other points in the color evolution were fitted with a fourth order polynomial with rms = 0.08.
\item PTF09cnd: The key epochs are available in the observed PTF09cnd light-curve. We fitted the epochs available for the $-10$d to $+40$d baseline  with a third order polynomial with rms = 0.04. The magnitudes from interpolation are in agreement with the direct measurements, and in Table~\ref{table:data} we report the actual magnitudes.
\item SCP-06F6:  The observed $z-$band was chosen as the filter to $K-$correct to retrieve the 400nm band rest frame, 
and the available data cover out to $\sim$20d past maximum.
 We interpolated the 6 epochs available between $-30$ days and 20 days from peak with a second and third order polynomial obtaining similar results. The best rms = 0.05 was found with the second order.  
As a consistency check, we also estimated what we would expect to see (assuming our fit is valid) in the $i$-band lightcurve at $\sim$40d, since an  $i-$band limit at this epoch is reported in \citet{ba09}. We find consistency and our 
$z$-band fit results would only be incompatible with the limit if the $i-$band mag would be about 1 mag fainter than 
the limit reported in \citet{ba09}. However no SLSN Ic has published data which show such a steep decrease or dramatic color change at this phase so far.
\item PTF11rks: The faintest object of the sample shows two small gaps of four and five days around 11d and 23d past maximum light. We fitted the 14 observations between $-3$d and +29d from peak  with a third order polynomial obtaining rms = 0.05. A polynomial of equivalent order fitted the color evolution on the same baseline with rms = 0.06. We noticed that magnitudes from interpolation of the fits extending up to +48 days are in agreement with those retrieved at +48d by SED fitting. The actual magnitude at peak was in agreement with that from interpolation.
\item PS1-10bzj: The data cover two out of four key epochs but thanks to a sufficient coverage (8 epochs) spanning over the baseline range defined above we retrieved the best fit with a polynomial of third order and rms = 0.07. 
We also fitted the color evolution on the same baseline with a third order polynomial retrieving rms = 0.06.
\item PS1-10pm: The data  (9 epochs) cover from $-4$d to +41d. We fitted them with a third order polynomial retrieving rms = 0.08.
\item LSQ12dlf: The key epochs are available in the observed LSQ12dlf light-curve. We fitted the 30 available epochs for our above-mentioned baseline (-10d to 40d) with a third order polynomial with 0.09 rms. The magnitudes from interpolation are in agreement with the direct measurements, and in Table~\ref{table:data} we report the actual magnitudes. The 12 epochs of the $M(400)-M(520)$ color evolution were fitted with a third order polynomial and 0.03 as rms.
\item SN2013dg:  We fitted 18 epochs from maximum to 42 days after with a third order polynomial (rms = 0.06). The same number of epochs were fitted for the color evolution retrieving an rms = 0.03.
\item PTF12dam: The lightcurve is not densely sampled at peak, but we are confident about the fit results and the magnitudes at key epochs because of the slowly evolving light curve. We fitted 9 epochs from $-15$d to +55d from peak with a third order polynomial and an rms = 0.06. The color evolution was also fitted in the same period with a third order polynomial and an rms = 0.04.
\item PS1-11ap: The data coverage is good with two small gaps around 13 and 25 days after peak. A third order polynomial was used to fit the 24 data points for $M(400)$ and  20 epochs of  color evolution $M(400)-M(520)$ in the range from $-16$d to +42d, retrieving rms = 0.10 and rms = 0.11, respectively. Despite the abundance of data at peak and +30d from that, we used the magnitudes from interpolation to be consistent through the sample.
\item CSS121015: We fitted the 24 epochs in the range $-15$d to +40d with a third order polynomial and rms = 0.09. Since the light curve is likely powered by interaction \citep{be14},  especially around peak, we decided to use the actual peak magnitude instead of that from the fit, while the other values ( in Table~\ref{table:data} ) come from the interpolation. We fit the color evolution in the same range of the light curve finding an rms = 0.10.
\end{itemize}

\begin{figure*}
\center
\includegraphics[width=18cm]{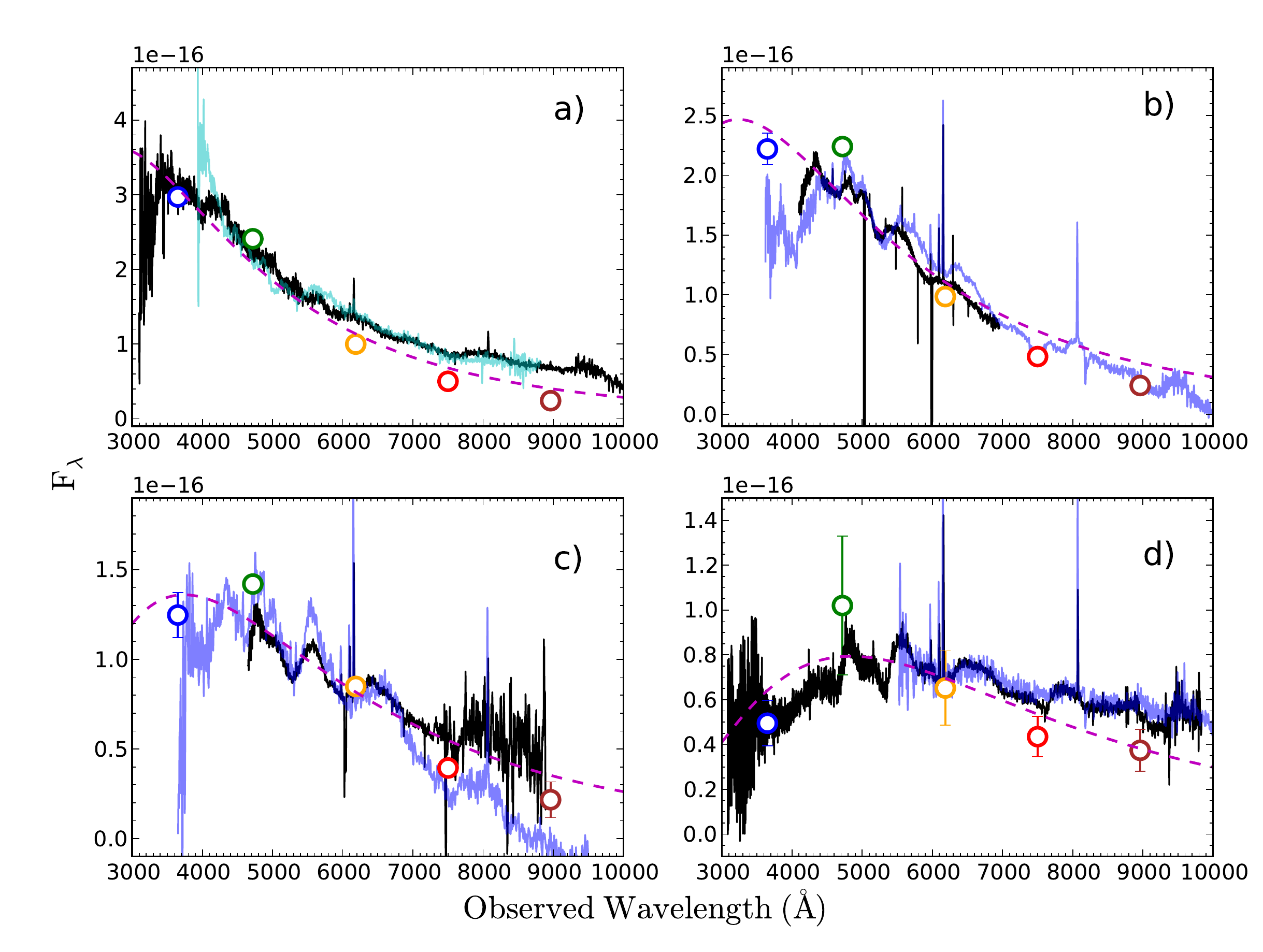}
\caption{{\em Panel a:} observed spectrum of SN2010gx around maximum light; the SED fit (in magenta) is a blackbody fit to the $ugriz$ photometry in blue, green, orange, red and brown, respectively; the template spectrum of SN2005ap at a similar epoch is plotted in cyan. {\em Panel b:} the same as Panel a but around +10d past maximum. The SN chosen and plotted as a template is SN2011ke (in blue). Panel c: same as panel b, but at +20d past maximum. Panel d: same as panel b, but at +30d past maximum.} 
\label{fig:Kcomp}
\end{figure*}

\section{Treatment of errors}\label{app:err}
When evaluating the errors on our absolute magnitudes, we tried as far as possible to account for all sources of error.  As reported in Sec.~\ref{ss:kc} our observed magnitude in the passband $f$ and absolute magnitude in the passband 400nm (the following discussion is valid also for the 520nm band) are related as follows :

\begin{equation}\label{eq:st}
m_{f} = M(400) + \mu  + K_{f \rightarrow 400} + A_{f}\: .
\end{equation}

According to the formalism reported in \citet{br07} the $K$-correction from $f$ to 400 is the following:

\begin{equation}
K_{f \rightarrow 400} = -2.5\: {\rm log}\left[ \frac{1}{1+z} \times \frac{\int d\lambda_{\rm o}\: \lambda_{\rm o}\: L_{\lambda}(\lambda_{\rm o}/1+z)\: f(\lambda_{\rm o}) \int d\lambda_{\rm e}\: \lambda_{\rm e}\: g_{\lambda}^{400}(\lambda_{\rm e})\: 400(\lambda_{\rm e})}{\int d\lambda_{\rm o}\: \lambda_{\rm o}\:  g_{\lambda}^{f}(\lambda_{\rm o})\: f(\lambda_{\rm o}) \int d\lambda_{\rm e}\: \lambda_{\rm e}\:L_{\lambda}(\lambda_{\rm e}) 400(\lambda_{\rm e})}\right]\ ,
\end{equation}

where L$_\lambda$ is the luminosity per unit wavelength,  $\lambda_{\rm o}$ refers to the observed frame, $\lambda_{\rm e}$ refers to the rest frame (e stands for  emitted), $g_{\lambda}^{400}$ is the flux density per unit wavelength for the 400nm band and $g_{\lambda}^{f}$ is the same for the observed band, while $400(\lambda)$ and $f(\lambda)$ are the response of the instrument per unit photon for the rest and observed filter, respectively. Rewriting equation \ref{eq:st}, we have :

\begin{equation}\label{eq:sti}
M(400)= m_{f} -(5\:{\rm log}\: D_{\rm L}+25)  - A_{f} - K_{f \rightarrow 400}\ ,
\end{equation}
where the $K$-correction term is not expanded and the distance modulus has been expressed as function of distance luminosity $D_{\rm L}$ in Mpc.
The errors on $M(400)$, nominally $\sigma_{M(400)}$ and equivalent to the standard deviation, are given by the combination of errors of the terms on the right hand side of equation \ref{eq:sti}. Assuming as first approximation that the terms are uncorrelated and the errors deriving from the cosmology adopted are negligible, the errors on our absolute magnitude follow the equation below:

\begin{equation}
\sigma_{\rm M(400)}= \sqrt{ \sigma^2_{\rm m_{\rm f}} + 4.715\left(\frac{\sigma_{\rm D_{\rm L}}}{D_{\rm L}}\right)^2  +\sigma^2_{\rm A_{\rm f}} + 1.179\left[ \left( \frac{\sigma_{z}}{z}\right)^2 + \left( \frac{\sigma_{L_{\lambda_{\rm o}}}}{L_{\lambda_{\rm o}}}\right)^2 + \left( \frac{\sigma_{L_{\lambda_{\rm e}}}}{L_{\lambda_{\rm e}}}\right)^2 + \left( \frac{\sigma_{ZP_{\lambda_{\rm o}}}}{ZP_{\lambda_{\rm o}}}\right)^2 + \left( \frac{\sigma_{ZP_{\lambda_{\rm e}}}}{ZP_{\lambda_{\rm e}}}\right)^2 \right] }\: 
\end{equation}

Where the differences between the Vega and AB systems are taken into account by the covariance of the zero points (ZP) of the two systems ($ZP_{\lambda_{\rm o}}$ and $ZP_{\lambda_{\rm e}}$). The two major sources of errors are related to observed magnitudes and extinction. When a cross-filter $K$-correction is made, the errors are usually in the range 0.005--0.05 mag \citep{1996PASP..108..190K,br07,2007ApJ...663.1187H}. These values are lower than, or comparable to, those of the 
errors on extinction and observed magnitudes. During our analysis we effectively always applied  a cross-filter $K$-correction. The errors related to the $K$-correction can be 3-4\% ($0.03-0.04$\,mag) if we use two different systems for the observed and rest filter band, as we had to do for SN2005ap, PTF09cnd, LSQ12dlf and CSS121015; and can be up to 0.05\,mag if the cross-filter $K$ correction is not applied \citep{2007ApJ...663.1187H}.

\begin{figure*}
\center
\includegraphics[width=18cm]{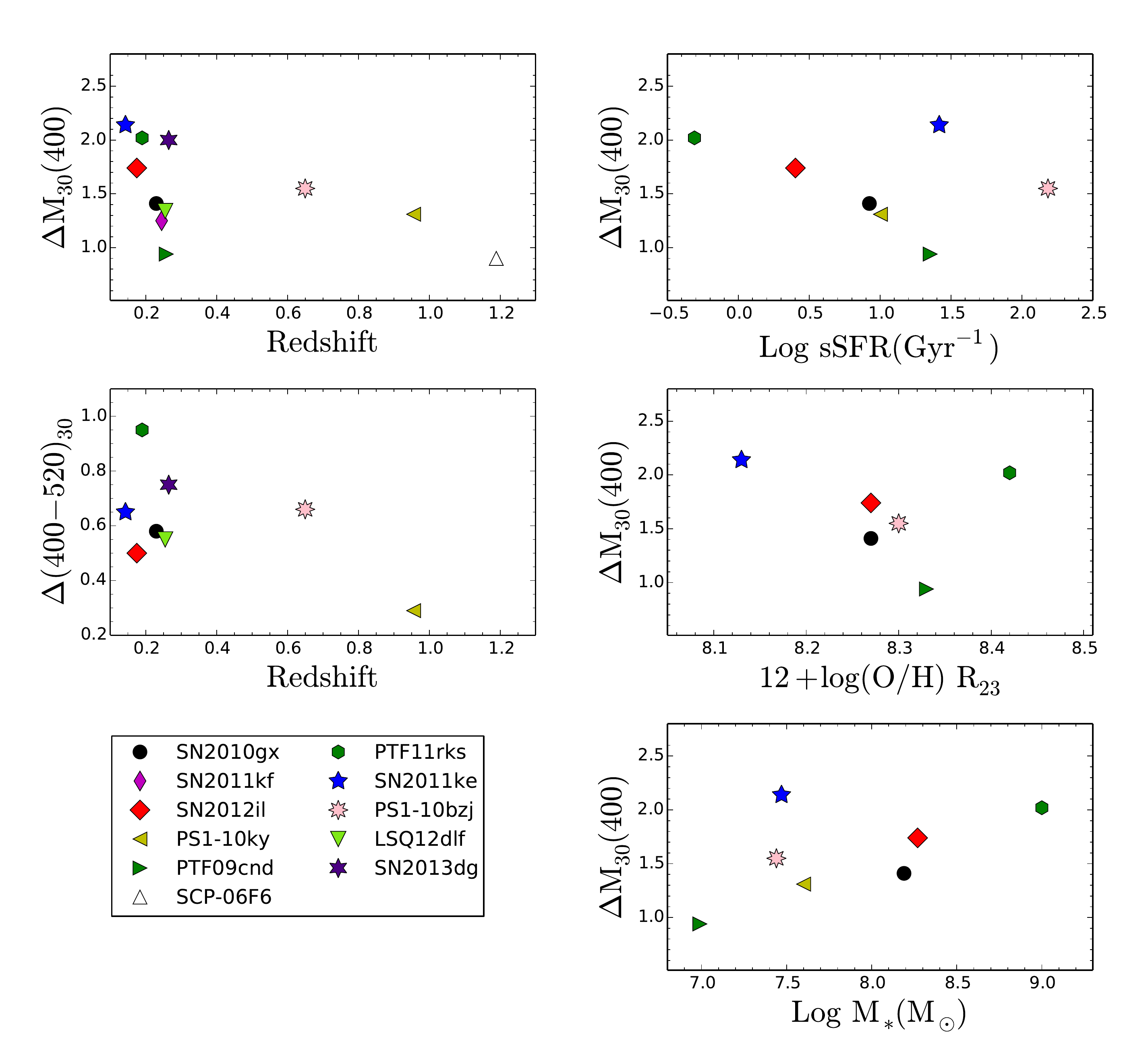}
\caption{Various plots of example empirical properties of SLSN derived in this paper compared to redshift and host galaxy properties from \citet{ch13,lu13b}.} 
\label{fig:prop}
\end{figure*}

The error treatment so far is applicable if we use the spectrum of the SN to evaluate the $K$-correction. If instead of the actual SN spectrum we use a library template spectrum (of a similar supernova) or the SED fitting (using blackbody spectra) to observed  photometry then some further scatter is found but $\sigma_ {M(400)}$ is not significantly affected by this. In order to quantitatively investigate  the differences  between the three methods, we compared the SED photometry fitting method  (employing blackbody fits) and the spectral library method (using our 64 spectra) as applied to SN2010gx. We compared these values to  those reported in Table~\ref{table:kcorr} which were evaluated using the actual SN spectra.  In Fig.~\ref{fig:Kcomp}  the three spectra used for the three methods are illustrated together. In comparison to the $K$-correction values derived with the SN spectra (reported in Table~\ref{table:kcorr} for SN2010gx),  the SED fitting method gives results which are different by  0.01--0.06 mag depending on the phase and the EW of the emission lines. The library template method results in differences 
which are between 0.01 to 0.05 mag, depending on the to filter and spectra used as template. We note that the major differences are found at the extremities of the optical wavelength region (e.g. $u$ and $z$). These potential systematic differences were not included in the final estimates of $\sigma_{M(400)}$ (since they are not strictly speaking random errors), but are given here to quantify the differences found  when using different methods for 
$K-$correction.

\section{The dependence of SLSN properties on their host galaxies}\label{app:prop}

In Figure~\ref{fig:prop} we illustrate an initial comparison of SLSN  properties as a function of redshift and host galaxy properties. Two of the empirical properties determined in this paper  ($\Delta M_{30}$(400) and $\Delta (400-520)_{30}$) are plotted against redshifts and  host galaxies properties from \citet{ch13} and \citet{lu13b}.  The decline parameter after 30 days
and the color evolution parameter are representative of the empirical properties we have investigated. 
The three galaxy properties used are specific star formation rate, total stellar mass and 
oxygen abundance. For oxygen abundance we took the $R_{23}$ estimate from \citet{lu13b}, using the low metallicity branch value (since the low masses of these SLSN hosts makes this branch choice most plausible).   As one might expect, from the small sample size, there is no  clear dependence of SLSN properties with host galaxy properties. The two plots comparing the decline parameters with 
redshift illustrate that the intrinsically fainter objects are not represented at 
the higher redshifts, which is not unexpected given the magnitude limited surveys that the discoveries
originate from. One might postulate a trend in the  $\Delta M_{30}$(400) vs $\log M_{\star}$ plot, 
in the sense that the intrinsically fainter SLSN tend to occur  in more massive galaxies. 
However this is possibly just driven by one point (either PTF09cnd at the low mass end or 
PTF11rks at the other mass extreme) and drawing conclusions would not be secure. 
Thus, an increase in the sample is mandatory in order to better investigate any underlying
physical relation or redshift evolution.


\begin{thebibliography}{}
\label{refs}
\bibitem[Agnoletto et al.(2009)]{2009ApJ...691.1348A} Agnoletto, I., 
Benetti, S., Cappellaro, E., et al.\ 2009, \apj, 691, 1348 
\bibitem[Baltay et al.(2013)]{2013PASP..125..683B} Baltay, C., Rabinowitz, 
D., Hadjiyska, E., et al.\ 2013, \pasp, 125, 683 
\bibitem[Barbary et al.(2009)]{ba09} Barbary, K., Dawson, 
K.~S., Tokita, K., et al.\ 2009, \apj, 690, 1358
\bibitem[Berger et al.(2012)]{be12} Berger, E., Chornock, 
R., Lunnan, R., et al.\ 2012, \apjl, 755, L29
\bibitem[\protect\citeauthoryear{Benetti et 
al.}{2014}]{be14} Benetti S., et al., 2014, MNRAS, 441, 289
\bibitem[Blanton 
\& Roweis(2007)]{br07} Blanton, M.~R., \& Roweis, S.\ 2007, \aj, 133, 734
\bibitem[Campana et al.(2006)]{ca06} Campana, S., Mangano, 
V., Blustin, A.~J., et al.\ 2006, \nat, 442, 1008
\bibitem[Cao et al.(2013)]{2013ApJ...775L...7C} Cao, Y., Kasliwal, M.~M., 
Arcavi, I., et al.\ 2013, \apjl, 775, L7 
\bibitem[Cano \& Jakobsson(2014)]{2014arXiv1409.3570C} Cano, Z., \& Jakobsson, P.\ 2014, arXiv:1409.3570 
\bibitem[Cano(2014)]{2014arXiv1407.2589C} Cano, Z.\ 2014, arXiv:1407.2589 
\bibitem[Chen et al.(2013)]{ch13} Chen, T.-W., Smartt, 
S.~J., Bresolin, F., et al.\ 2013, \apjl, 763, L28 
\bibitem[Chevalier 
\& Irwin(2011)]{2011ApJ...729L...6C} Chevalier, R.~A., \& Irwin, C.~M.\ 2011, \apjl, 729, L6 
\bibitem[Chomiuk et al.(2011)]{ch11} Chomiuk, L., Chornock, 
R., Soderberg, A.~M., et al.\ 2011, \apj, 743, 114 
\bibitem[Chornock et al.(2013)]{chor13} Chornock, R., Berger, 
E., Rest, A., et al.\ 2013, \apj, 767, 162
\bibitem[\protect\citeauthoryear{Cobb et al.}{2006}]{co06} 
Cobb B.~E., Bailyn C.~D., van Dokkum P.~G., Natarajan P., 2006, ApJ, 645, 
L113
\bibitem[Cooke et al.(2012)]{co12} Cooke, J., Sullivan, M., 
Gal-Yam, A., et al.\ 2012, \nat, 491, 228
\bibitem[D'Andrea et al.(2010)]{2010ApJ...708..661D} D'Andrea, C.~B., Sako, 
M., Dilday, B., et al.\ 2010, \apj, 708, 661 
\bibitem[Dessart et al.(2012)]{de12} Dessart, L., Hillier, 
D.~J., Waldman, R., Livne, E., \& Blondin, S.\ 2012, \mnras, 426, L7
\bibitem[Drake et al.(2011)]{dr11} Drake, A.~J., Djorgovski, 
S.~G., Mahabal, A.~A., et al.\ 2011, The Astronomer's Telegram, 3343, 1
\bibitem[Drake et al.(2009)]{dr09} Drake, A.~J., Djorgovski, 
S.~G., Mahabal, A., et al.\ 2009, \apj, 696, 870
\bibitem[Dunkley et al.(2009)]{dun09} Dunkley, J., Komatsu, 
E., Nolta, M.~R., et al.\ 2009, \apjs, 180, 306
\bibitem[Eisenstein et al.(2005)]{eis05} Eisenstein, D.~J., 
Zehavi, I., Hogg, D.~W., et al.\ 2005, \apj, 633, 560 
\bibitem[Foley et al.(2003)]{fo03} Foley, R.~J., Papenkova, 
M.~S., Swift, B.~J., et al.\ 2003, \pasp, 115, 1220
\bibitem[Galama et al.(1998)]{ga98} Galama, T.~J., 
Vreeswijk, P.~M., van Paradijs, J., et al.\ 1998, \nat, 395, 670
\bibitem[Gal-Yam(2012)]{gy12} Gal-Yam, A.\ 2012, Science, 
337, 927
\bibitem[Gal-Yam et al.(2009)]{gy09} Gal-Yam, A., Mazzali, 
P., Ofek, E.~O., et al.\ 2009, \nat, 462, 624
\bibitem[Gehrels(1986)]{1986ApJ...303..336G} Gehrels, N.\ 1986, \apj, 303, 336 
\bibitem[Ginzburg 
\& Balberg(2012)]{2012ApJ...757..178G} Ginzburg, S., \& Balberg, S.\ 2012, \apj, 757, 178 
\bibitem[\protect\citeauthoryear{Goldhaber et 
al.}{2001}]{go01} Goldhaber G., et al., 2001, ApJ, 558, 359 
\bibitem[Guy et 
al.(2007)]{guy07} Guy, J., Astier, P., Baumont, S., et al.\ 2007, \aap, 466, 11 
\bibitem[Guy et 
al.(2005)]{guy05} Guy, J., Astier, P., Nobili, S., Regnault, N., \& Pain, R.\ 2005, \aap, 443, 781 
\bibitem[Hamuy 
\& Pinto(2002)]{2002ApJ...566L..63H} Hamuy, M., \& Pinto, P.~A.\ 2002, \apjl, 566, L63 
\bibitem[Hamuy et al.(1996)]{ham96} Hamuy, M., Phillips, 
M.~M., Suntzeff, N.~B., et al.\ 1996, \aj, 112, 2438
\bibitem[Hamuy et al.(1993)]{1993PASP..105..787H} Hamuy, M., Phillips, 
M.~M., Wells, L.~A., \& Maza, J.\ 1993, \pasp, 105, 787 
\bibitem[Hogg et al.(2002)]{2002astro.ph.10394H} Hogg, D.~W., Baldry, 
I.~K., Blanton, M.~R., \& Eisenstein, D.~J.\ 2002,
arXiv:astro-ph/0210394 
\bibitem[Howell et al.(2013)]{ho13} Howell, D.~A., Kasen, 
D., Lidman, C., et al.\ 2013, \apj, 779, 98
\bibitem[Hsiao et al.(2007)]{2007ApJ...663.1187H} Hsiao, E.~Y., Conley, A., 
Howell, D.~A., et al.\ 2007, \apj, 663, 1187 
\bibitem[Hunter et 
al.(2009)]{hu09} Hunter, D.~J., Valenti, S., Kotak, R., et al.\ 2009, \aap, 508, 371 
\bibitem[Jones et al.(2013)]{2013ApJ...768..166J} Jones, D.~O., Rodney, 
S.~A., Riess, A.~G., et al.\ 2013, \apj, 768, 166 
\bibitem[Kaiser et al.(2010)]{PS1_system} Kaiser, N., et al. \ 2010, \procspie, 7733,  12K.
\bibitem[Kasen 
\& Bildsten(2010)]{kb10} Kasen, D., \& Bildsten, L.\ 2010, \apj, 717, 245 
\bibitem[Kim et al.(1996)]{1996PASP..108..190K} Kim, A., Goobar, A., 
\& Perlmutter, S.\ 1996, \pasp, 108, 190 
\bibitem[Komatsu et al.(2011)]{kom11} Komatsu, E., Smith, 
K.~M., Dunkley, J., et al.\ 2011, \apjs, 192, 18 
\bibitem[Kostrzewa-Rutkowska et al.(2013)]{ko13} 
Kostrzewa-Rutkowska, Z., Koz{\l}owski, S., Wyrzykowski, {\L}., et al.\ 
2013, \apj, 778, 168 
\bibitem[King et al.(2014)]{ki13} King, A.~L., Davis, T.~M., 
Denney, K.~D., Vestergaard, M., \& Watson, D.\ 2014, \mnras, 441, 3454
\bibitem[Inserra et al.(2013)]{in13} Inserra, C., Smartt, 
S.~J., Jerkstrand, A., et al.\ 2013, \apj, 770, 128
\bibitem[Inserra et al.(2012)]{in12a} Inserra, C., Smartt, 
S.~J., Fraser, M., et al.\ 2012, The Astronomer's Telegram, 4329, 1
\bibitem[Laureijs et al.(2011)]{2011arXiv1110.3193L} Laureijs, R., Amiaux, 
J., Arduini, S., et al.\ 2011, arXiv:1110.3193
\bibitem[Leloudas et 
al.(2012)]{le12} Leloudas, G., Chatzopoulos, E., Dilday, B., et al.\ 2012, \aap, 541, A129 
\bibitem[Lord(1992)]{lo92} Lord, S.~D.\ 1992, NASA Technical Memorandum 103957
\bibitem[Lunnan et al.(2014)]{lu13b} Lunnan, R., Chornock, 
R., Berger, E., et al.\ 2014, \apj, 787, 138
\bibitem[Lunnan et al.(2013)]{lu13} Lunnan, R., Chornock, 
R., Berger, E., et al.\ 2013a, \apj, 771, 97
\bibitem[Maguire et al.(2010)]{2010MNRAS.403L..11M} Maguire, K., Kotak, R., 
Smartt, S.~J., et al.\ 2010, \mnras, 403, L11 
\bibitem[Mandel et al.(2011)]{man11} Mandel, K.~S., Narayan, 
G., \& Kirshner, R.~P.\ 2011, \apj, 731, 120
\bibitem[Mandel et al.(2009)]{man09} Mandel, K.~S., 
Wood-Vasey, W.~M., Friedman, A.~S., 
\& Kirshner, R.~P.\ 2009, \apj, 704, 629
\bibitem[Mazzali et al.(2008)]{ma08} Mazzali, P.~A., 
Valenti, S., Della Valle, M., et al.\ 2008, Science, 321, 1185
\bibitem[McCrum et al.(2014b)]{mc14b} McCrum, M., Smartt, 
S.~J., Rest, A., et al.\ 2014, arXiv:1402.1631
\bibitem[McCrum et al.(2014a)]{mc13} McCrum, M., Smartt, 
S.~J., Kotak, R., et al.\ 2014, \mnras, 2595
\bibitem[McKenzie 
\& Schaefer(1999)]{mck99} McKenzie, E.~H., \& Schaefer, B.~E.\ 1999, \pasp, 111, 964
\bibitem[Mirabal et al.(2006)]{mi06} Mirabal, N., Halpern, 
J.~P., An, D., Thorstensen, J.~R., 
\& Terndrup, D.~M.\ 2006, \apjl, 643, L99 
\bibitem[Modjaz et al.(2009)]{mo09} Modjaz, M., Li, W., 
Butler, N., et al.\ 2009, \apj, 702, 226
\bibitem[Nicholl et al.(2013)]{ni13} Nicholl, M., Smartt, 
S.~J., Jerkstrand, A., et al.\ 2013, \nat, 502, 346
\bibitem[Nicholl et al.(2014)]{ni14} Nicholl, M., Smartt, 
S.~J., Jerkstrand, A., et al.\ 2014, \mnras, 444, 2096
\bibitem[Nugent et al.(2006)]{2006ApJ...645..841N} Nugent, P., Sullivan, 
M., Ellis, R., et al.\ 2006, \apj, 645, 841 
\bibitem[Nugent et al.(2002)]{nu02} Nugent, P., Kim, A., 
\& Perlmutter, S.\ 2002, \pasp, 114, 803
\bibitem[Oke 
\& Sandage(1968)]{1968ApJ...154...21O} Oke, J.~B., \& Sandage, A.\ 1968, \apj, 154, 2
\bibitem[Pandey et al.(2003)]{pa03} Pandey, S.~B., Anupama, 
G.~C., Sagar, R., et al.\ 2003, \mnras, 340, 375
\bibitem[Pastorello et al.(2010a)]{pa10} Pastorello, A., 
Smartt, S.~J., Botticella, M.~T., et al.\ 2010, \apjl, 724, L16 
\bibitem[Pastorello et al.(2010b)]{pa10b} Pastorello, A., 
Smartt, S.~J., Kankare, D., et al.\ 2010, The Astronomer's Telegram, 2504, 
1
\bibitem[Pastorello et al.(2008)]{pa08} Pastorello, A., 
Kasliwal, M.~M., Crockett, R.~M., et al.\ 2008, \mnras, 389, 955
\bibitem[Patat et al.(2001)]{pat01} Patat, F., Cappellaro, 
E., Danziger, J., et al.\ 2001, \apj, 555, 900
\bibitem[Patat et al.(2007)]{2007Sci...317..924P} Patat, F., Chandra, P., 
Chevalier, R., et al.\ 2007, Science, 317, 924 
\bibitem[Percival et al.(2010)]{per10} Percival, W.~J., Reid, 
B.~A., Eisenstein, D.~J., et al.\ 2010, \mnras, 401, 2148
\bibitem[Perlmutter et al.(1997)]{pe97} Perlmutter, S., 
Gabi, S., Goldhaber, G., et al.\ 1997, \apj, 483, 565
\bibitem[Phillips(1993)]{ph93} Phillips, M.~M.\ 1993, \apjl, 
413, L105 
\bibitem[Pian et al.(2006)]{pi06} Pian, E., Mazzali, P.~A., 
Masetti, N., et al.\ 2006, \nat, 442, 1011
\bibitem[Poznanski et al.(2009)]{2009ApJ...694.1067P} Poznanski, D., 
Butler, N., Filippenko, A.~V., et al.\ 2009, \apj, 694, 1067 
\bibitem[Prieto et al.(2012)]{pr12} Prieto, J.~L., Drake, 
A.~J., Mahabal, A.~A., et al.\ 2012, The Astronomer's Telegram, 3883, 1
\bibitem[Prieto et al.(2006)]{pr06} Prieto, J.~L., Rest, A., 
\& Suntzeff, N.~B.\ 2006, \apj, 647, 501
\bibitem[Pskovskii(1977)]{ps77} Pskovskii, I.~P.\ 1977, 
\sovast, 21, 675
\bibitem[Quimby et al.(2014)]{qu14} Quimby, R.~M., Oguri, 
M., More, A., et al.\ 2014, Science, 344, 396
\bibitem[Quimby et al.(2013b)]{qu13b} Quimby, R.~M., Yuan, F., 
Akerlof, C., \& Wheeler, J.~C.\ 2013, \mnras, 431, 912 
\bibitem[Quimby et al.(2013a)]{qu13a} Quimby, R.~M., Werner, 
M.~C., Oguri, M., et al.\ 2013, \apjl, 768, L20
\bibitem[Quimby et al.(2011a)]{qu11} Quimby, R.~M., Kulkarni, 
S.~R., Kasliwal, M.~M., et al.\ 2011, \nat, 474, 487
\bibitem[Quimby et al.(2011b)]{qu11b} Quimby, R.~M., Gal-Yam, 
A., Arcavi, I., et al.\ 2011, The Astronomer's Telegram, 3841, 1 
\bibitem[Quimby et al.(2011c)]{qu11c} Quimby, R.~M., 
Sternberg, A., \& Matheson, T.\ 2011, The Astronomer's Telegram, 3344, 1
\bibitem[Quimby et al.(2010)]{qu10} Quimby, R.~M., Kulkarni, 
S.~R., Ofek, E., et al.\ 2010, The Astronomer's Telegram, 2492, 1 
\bibitem[Quimby et al.(2007)]{qu07} Quimby, R.~M., Aldering, 
G., Wheeler, J.~C., et al.\ 2007, \apjl, 668, L99
\bibitem[Quimby(2006)]{2006CBET..644....1Q} Quimby, R.\ 2006, Central 
Bureau Electronic Telegrams, 644, 1 
\bibitem[Quimby et al.(2005)]{qu05} Quimby, R.~M., Castro, 
F., Gerardy, C.~L., et al.\ 2005, Bulletin of the American Astronomical 
Society, 37, \#171.02
\bibitem[Rau et al.(2009)]{rau09} Rau, A., Kulkarni, S.~R., 
Law, N.~M., et al.\ 2009, \pasp, 121, 1334 
\bibitem[Richmond et al.(1996)]{ri96} Richmond, M.~W., van 
Dyk, S.~D., Ho, W., et al.\ 1996, \aj, 111, 327
\bibitem[Riess et al.(1998)]{rie98} Riess, A.~G., Filippenko, 
A.~V., Challis, P., et al.\ 1998, \aj, 116, 1009 
\bibitem[Riess et al.(1996)]{rie96} Riess, A.~G., Press, 
W.~H., \& Kirshner, R.~P.\ 1996, \apj, 473, 88
\bibitem[Rodney et al.(2012)]{2012ApJ...746....5R} Rodney, S.~A., Riess, 
A.~G., Dahlen, T., et al.\ 2012, \apj, 746, 5 
\bibitem[Rust(1974)]{ru74} Rust, B.~W.\ 1974, \baas, 6, 309
\bibitem[Smith et al.(2007)]{2007ApJ...666.1116S} Smith, N., Li, W., Foley, 
R.~J., et al.\ 2007, \apj, 666, 1116 
\bibitem[Smith 
\& McCray(2007)]{2007ApJ...671L..17S} Smith, N., \& McCray, R.\ 2007, \apjl, 671, L17
\bibitem[Soderberg et al.(2008)]{sod08} Soderberg, A.~M., 
Berger, E., Page, K.~L., et al.\ 2008, \nat, 453, 469 
\bibitem[Sollerman et 
al.(2006)]{so06} Sollerman, J., Jaunsen, A.~O., Fynbo, J.~P.~U., et al.\ 2006, \aap, 454, 503
\bibitem[Sollerman et al.(2000)]{so00} Sollerman, J., Kozma, 
C., Fransson, C., et al.\ 2000, \apjl, 537, L127
\bibitem[Stritzinger et al.(2009)]{st09} Stritzinger, M., 
Mazzali, P., Phillips, M.~M., et al.\ 2009, \apj, 696, 713
\bibitem[Suzuki et al.(2012)]{su12} Suzuki, N., Rubin, D., 
Lidman, C., et al.\ 2012, \apj, 746, 85 
\bibitem[Tanaka et al.(2012)]{2012MNRAS.422.2675T} Tanaka, M., Moriya, 
T.~J., Yoshida, N., \& Nomoto, K.\ 2012, \mnras, 422, 2675 
\bibitem[Tomita et al.(2006)]{to06} Tomita, H., Deng, J., 
Maeda, K., et al.\ 2006, \apj, 644, 400
\bibitem[Tonry et al.(2012)]{2012ApJ...750...99T} Tonry, J.~L., Stubbs, 
C.~W., Lykke, K.~R., et al.\ 2012, \apj, 750, 99 
\bibitem[Valenti et al.(2011)]{va11} Valenti, S., Fraser, 
M., Benetti, S., et al.\ 2011, \mnras, 416, 3138 
\bibitem[Valenti et al.(2008a)]{va08a} Valenti, S., Benetti, 
S., Cappellaro, E., et al.\ 2008, \mnras, 383, 1485 
\bibitem[Valenti et al.(2008b)]{va08b} Valenti, S., 
Elias-Rosa, N., Taubenberger, S., et al.\ 2008, \apjl, 673, L155
\bibitem[Woosley(2010)]{2010ApJ...719L.204W} Woosley, S.~E.\ 2010, \apjl, 
719, L204 
\bibitem[Yoshii et al.(2003)]{yo03} Yoshii, Y., Tomita, H., 
Kobayashi, Y., et al.\ 2003, \apj, 592, 467
\bibitem[Young et 
al.(2010)]{yo10} Young, D.~R., Smartt, S.~J., Valenti, S., et al.\ 2010, \aap, 512, A70
\end{thebibliography}
\end{document}